\documentclass[a4]{article}

\usepackage{natbib}
\usepackage{graphicx}
\usepackage{color} 
\usepackage{verbatim,framed}
\usepackage{algorithm2e}

\definecolor{orange}{rgb}{1.0,0.5,0.0}
\definecolor{aqgr}  {rgb}{0.0,1.0,0.6} % aqua green
\definecolor{viol}  {rgb}{0.8,0.6,1.0}

\title{Epigenetic Tracking: Implementation Details}

\author{Alessandro Fontana$^{1}$ \\
\mbox{}\\
$^1$IEEE \\
alessandro.fontana@ieee.org}

\begin{document}
\maketitle

\begin{abstract}
``Epigenetic Tracking'' is the name of a model of cellular development that, coupled with an evolutionary technique, becomes an evo-devo (or ``artificial embryology'', or ``computational development'') method to generate 2d or 3d sets of artificial cells arbitrarily shaped. 'In silico' experiments have proved the effectiveness of the method in devo-evolving any kind of shape, of any complexity (in terms of number of cells, number of colours, etc.); being shape complexity a metaphor for organismal complexity, such simulations established its potential to generate the complexity typical of biological systems. Moreover, it has also been shown how the underlying model of cellular development is able to produce the artificial version of key biological phenomena such as embryogenesis, the presence of ``junk DNA'', the phenomenon of ageing and the process of carcinogenesis. The objective of this document is not to provide new material (most of the material presented here has already been published elsewhere): rather, it is to provide all details that, for lack of space, could not be provided in the published papers and in particular to give all technical details necessary to re-implement the method.
\end{abstract}

\section{The model of development}

\subsection{Introduction}

%\colorbox{green}{AE related work}
The previous work in the field of Artificial Embryology (see \citep{AY03SM} for a comprehensive review) can be divided into two broad categories: the grammatical approach and the cell chemistry approach. In the grammatical approach development is guided by sets of grammatical rewrite rules; context-free or context-sensitive grammars, instruction trees or directed graphs can be used; L-systems were first introduced by Lindenmayer \citep{AY68LX} to describe the complex fractal patterns observed in the structure of trees. The cell chemistry approach draws inspiration from the early work of Turing \citep{AY52TX}, who introduced reaction and diffusion equations to explain the striped patterns observed in nature (e.g. shells and animals' fur); this approach attempts to simulate cell biology at a deeper level, going inside cells and reconstructing the dynamics of chemical reactions and the networks of chemical signals exchanged between cells. Notable examples of grammatical embryogenies are \citep{AY68LX}, \citep{AY02HP}, \citep{AY94CN} and \citep{AY96GW}; among cell chemistry embryogenies, we recall \citep{AY69KF} and, more recently, \citep{AY01BP} and \citep{AY03MB}.

%\colorbox{green}{E.T. intro}
``Epigenetic Tracking'' (E.T.), first described in \citep{AY08AX}, is the name of a model of cellular development that, coupled with an evolutionary technique, becomes an evo-devo (or ``artificial embryology'', or ``computational development'') method to generate 2d or 3d sets of artificial cells arbitrarily shaped. The method evolves instructions contained in the cells' genome, which guide the development of an artificial organism from a set of zygotes. 'In silico' experiments have proved the effectiveness of the method in devo-evolving any kind of shape, of any complexity (in terms of number of cells, number of colours, etc.); being shape complexity a metaphor for organismal complexity, such simulations established its potential to generate the complexity typical of biological systems. Furthermore, it has also been shown \citep{AY09AX} how the underlying model of development is able to produce the artificial version of key biological phenomena such as embryogenesis, the presence of ``junk DNA'', the phenomenon of ageing and the process of carcinogenesis. The objective of this document is to provide all technical details necessary to re-implement the method. The document is organised as follows: section 1 describes in detail the model of development; section 2 describes evo-devo method; sections 3 reports parameters and pseudocode of the algorithm (all figures are at the end of the document).

\subsection{Normal cells, driver cells, views}

%\colorbox{green}{normal and driver cells}
Artificial organisms are represented as sets of square- or cube-shaped artificial cells deployed on a grid (see figure 1). Development starts with one (or more) cell(s) -called zygote(s)- placed on the grid and unfolds in ASMAX steps, counted by the variable 'age step' (AS), which is shared by all cells and can be considered the ``global clock'' of the organism. Cells belong to two categories: 'normal cells' (which make up the bulk of the shape) and 'driver cells' (which are much fewer in number and are uniformly spread throughout the shape). Driver cells have an associated Genome (organised as an array of 'change instructions') and a variable called 'cellular epigenetic type' (CET, organised as an array of ASMAX integers).

%\colorbox{green}{differentiation and views}
While the Genome is identical for all driver cells, the CET value is different in each driver cell: in this way, it can be used by different driver cells as a ``key'' to activate different instructions in the Genome (see figure 2). The CET value represents the source of cell differentiation during development, allowing driver cells to behave differently despite sharing the same Genome. A shape can be ``viewed'' in two ways (see figure 3). In ``external view'' (right in the figure) cells are shown with their colours; in ``internal view'' (left in the figure) colours represent cell properties: blue is used for normal cells alive, orange for normal cells just (i.e. in the current age step) created, grey for cells that have just died, yellow for driver cells (regardless of when they have been created). The model of development can be applied to 2d and 3d organisms: from now on, for generality reasons, we will make reference to the 3d case, except for the figures which, for simplicity reasons, will mostly refer to the 2d case.

\subsection{Structure of CET values}

%\colorbox{green}{structure of CET values: 0}
The CET values are arrays of size ASMAX (the total number of age steps foreseen); ASMAX includes also the 0th step (in which only the zygote(s) is (are) present on the grid and nothing else happens): therefore the values of the steps are: 0,1,2,...ASMAX-1. The first position of the array contains a number that identifies the zygote; in case more zygotes are present (a situation theoretically possible, although biologically not plausible), the first zygote's CET value will have a '0' in the first position, the second zygote's CET value will have a '1', the third zygote's CET value will have a '2' and so forth; in zygotes all other array positions are filled with '0' (e.g. if ASMAX=4, the first zygote's CET value is [0,0,0,0], the second zygote's CET value is [1,0,0,0], the third zygote's CET value is [2,0,0,0], etc.), while in other driver cells the other array positions contain numbers reflecting the driver's generative history.

%\colorbox{green}{structure of CET values: 1}
In the previous definition the parameter ASMAX (total number of age steps) has been considered known and fixed at the beginning of development; actually this assumption is not strictly necessary: development could start with a value ASMAX0 for ASMAX, that could be subsequently increased to another value ASMAX1 greater than ASMAX0; in this case the positions from ASMAX0 to ASMAX1-1 of the CET arrays generated in the first ASMAX0 steps would be automatically filled with '0'. Alternatively, the size of the CET values could become variable, incremented by one unit at each age step (e.g. the zygote's CET value would be [0], the driver cells' CET values created in step 1 would be [0,1],[0,2],[0,3],..., the driver cells' CET values created in step 2 would be [0,2,1],[0,2,2],[0,2,3], etc. In the remainder of this document we will assume for simplicity reasons the presence of a single zygote and a fixed value for ASMAX.

\subsection{Structure of instructions}

%\colorbox{green}{genome}
The Genome is organised as an array of change instructions (see figure 4), having a left part and a right part. The left part comprises the following fields:

\begin{itemize}
\item a field called ON (integer, possible values: 0,1). It tells whether the instruction is active (1) or inactive (0);
\item a field called OP ('order of precedence', integer, possible values: [0-maxval]). It is used to decide which instruction is activated in case of conflict;
\item a field called XS (integer, possible values: [0-maxval] or -1). It contains the AS value at which the instruction can be activated (if XS=-1, the instruction can be activated at any step); 
\item a field called XET (array of integers, of the same size as CET, each with possible values: [0-maxval]). It contains the CET value that cna trigger the instruction's activation.
\end{itemize}

The right part of the instruction contains the code for the operations which are carried out when the instruction becomes active, which result in the creation and deletion of cells. The instruction's right part comprises the following fields:

\begin{itemize}
\item a field called ETP ('event type', integer, possible values: 0,1). It defines the type of 'change event' coded by the instruction: proliferation (value=0) or apoptosis (value=1);
\item a field called PC ('parallelepiped corners', array of 6 integers, each with possible values: [0-maxval]). It defines the 'change parallelepiped' (the parallelepiped which inscribes the 'temporary change volume') by specifying the coordinates of its north-west-back corner (first 3 integers) and south-east-front corner (last 3 integers);
\item a field called RM ('rotation matrix', array of 9 integers, possible values: [0-maxval]). It defines the rotation matrix to be applied to the temporary change volume to get the (final) change volume; 
\item a field called COL ('colour', integer, possible values: [0-maxval]). It specifies the colour of the cells created (in case of proliferation).
\end{itemize}

\subsection{Change events and development}

%\colorbox{green}{(CET,XET) match}
As we said, development unfolds in ASMAX age steps. At each step, for each driver cell and for each active instruction (ON $\neq$ 0), the algorithm tests if the instruction's XET matches the driver's CET and if the instruction's XS matches AS (this latter test is carried out only if XS is not -1, in which case we speak of ``timed'' instructions): if both matches are verified, it triggers the execution of the instruction's right part. If a CET value matches multiple XET values (CET values are guaranteed to be unique, XET values are not), the instruction with the highest OP value is executed; in case of multiple (CET,XET) matches, instructions matching CET values belonging to drivers with lower sums of coordinate values are executed first; a parameter of the algorithm (CGEVMAX) specifies the maximum number of instructions which can ``fire'' in a single step. The driver cells that triggered the instruction's execution, called 'mother cell', is always removed from the grid (otherwise the instruction would be activated over and over again). The pseudocode for the basic development loop is contained in subroutine Devloop, reported in the pseudocode section. 

%\colorbox{green}{right part}
The (CET,XET) and (AS,XS) matches cause the execution of the instruction's right part, which determines the occurrence of a 'change event', characterised by three things: type, shape and colour. Instructions are of two 'types': 'proliferation instructions' cause the matching driver cell to proliferate in a volume called 'change volume', 'apoptosis instructions' cause cells in the change volume to be deleted from the grid; the parameter 'shape' specifies the shape of the change volume, in which the proliferation/apoptosis events occur; in case of proliferation, the parameter 'colour' specifies the colour of the new cells. We note that the first event of development is necessarily a proliferation event triggered on the zygote: the other possibility would be an apoptosis event triggered on the zygote, but this would simply remove the zygote from the grid without creating any other cell and development would stop: a quite uninteresting behaviour. Figures 6, 7 and 8 show examples of proliferation from the zygote, proliferation from a generic driver cell and apoptosis respectively.

%\colorbox{green}{generalisations}
\color{blue} {\scshape Possible generalisations.} i) Conflict-resolution rules. The conflict-resolution rules described above, used to decide which instructions are to be activated in case of mutliple matches, appear (and are) rather arbitrary: alternative rules (e.g. executing first instructions triggered on drivers with lower sums of coordinate values) do not cause the system's behaviour to be significantly different, as long as the rule is deterministic. ii) Other types of change events. Other types of change events are possible, e.g. change events that combine proliferation and apoptosis or change events that 'colour' subsets of cells previously marked (all cells generated in a proliferation event could be assigned a code that identifies them for subsequent actions). More in general, change events can be defined as actions that change some cell properties for a given subset of cells. \color{black}

%\colorbox{green}{generalisations}
\color{blue} iii) Match. The simple (CET,XET) match, implemented by a function which returns 'TRUE' iff CET=XET, is only a possibility among many. More generally, the right part's activation can be bound to a generic function F(CET,XET) taking the value 'TRUE'. Possible examples of F are: *) a function F0 which returns 'TRUE' iff a number N $>$ ASMAX of array positions are equal (e.g. only the 3rd, the 5th and the 11th positions need to be equal for F0 to return 'TRUE'); *) a function F1 which returns 'TRUE' iff sub-arrays of length X within CET and XET are equal (e.g. only the sub-arrays from the 4th to the 10th positions need to be equal for F1 to return 'TRUE'). These two examples of F represent possibile generalisations of the simple 'match' function employed in the current implementation; in this way the activation of instructions would be less specific and could occur several times during development and in several shape 'loci', allowing code-reuse and possibly leading to more regularity in development. \color{black}

\subsection{Shaping primitives}

%\colorbox{green}{definition of change volume}
The 'change volume' is the volume (i.e. the subset of grid points) in which the events of proliferation and apoptosis occur. In practice this means that, in a proliferation event, the change volume is filled with new cells (both normal and driver) while, in case of an apoptosis event, all cells present in the change volume 'die', i.e. they are removed from the grid. The position of the change volume is always defined reletive to the mother cell's position such that, if the mother cell is moved, the change volume is moved with it; the mother cell's position can be either internal or external to the change volume (being the internal case the standard one). The shape of the change volume is selected from a set of basic shapes called 'shaping primitives', whose definition is achieved through a parametric approach; in such approach the primitive is defined through a set of numbers, grouped in two fields (belonging to the instruction's right part). The 'PC' (parallelepiped corners) field comprises 3+3=6 numbers that give the coordinates of the north-west-back and south-east-front corners of the so-called 'change parallelepiped', the 'RM' field comprises 9 numbers defining the 'rotation matrix'. The definition of the change volume is carried out in three processing steps (figures report examples for the 2d case, 3d extension is straightforward):

\begin{enumerate}
\item the 'change parallelepiped' is determined with the PC field (knowing the two diagonally opposite corners, the parallelepiped can be recostructed completely - figure 5-a1);
\item the 'temporary change volume' is defined as the set of all points contained in the ellissoid inscribed in the change parallelepiped (figure 5-a2);  
\item the final change volume is defined by applying the rotation defined by the 'rotation matrix' to the temporary change volume (figure 5-a3).
\end{enumerate}

%\colorbox{green}{generalisations}
\color{blue} {\scshape Possible generalisations.} The definition of shaping primitives can be achieved in several ways: the most straightforward one consists in providing the list of all available predefined shapes (see figure 5-b); in this case the right part of the instruction specifies the primitive to be used simply with the number representing the position of the primitive in the list. Compared to this simple list approach, the approach implemented has the advantage of allowing the generation of a much greater number of primitives: with 6+9=15 parameters, each with, say, 10 possible values, as many as $10^{15}$ different primitives can be generated, allowing a much higher degree of flexibility. The application of the ellissoid has been introduced to generate more ``rounded'' primitives, which have a more natural aspect. Different kinds of parametric approaches, as well as mixed list-parametric approaches, are possible and have been experimented with good results in other implementation attempts. Bases on this experience, we can conclude that the definition of the primitive set is not particularly critical for the algorithm functioning, unless of course it is not too ill-defined (e.g. a set containing only pyramid-shaped primitives to develop a round shape). \color{black}
  
\subsection{Deployment of driver cells, generation of CET values (proliferation)}

%\colorbox{green}{deployment of driver cells}
During a proliferation event, as we said, normal and driver cells are deployed on the change volume; normal cells fill the whole change volume, driver cells are much fewer and are ``sprinkled'' on the change volume. In the current implementation, a simple scan of the change volume is performed and driver cells are placed at regular intervals (e.g. one driver every NDRAT normal -NDRAT is a parameter of the algorithm) in each dimension: with this rule the ratio between normal cells and driver cells is $NDRAT^{3}$. Here is the pseudocode of the procedure ((x0,y0,z0) and (x1,y1,z1) are the parallelepiped corners, the procedure is applied before roto-translation):

\begin{verbatim}
For ax = x0 To = x1
   For ay = y0 To = y1
      For az = z0 To = z1
         If [(ax,ay,az) lies inside ellissoid]
         If (ax mod NDRAT = 0) and (ay mod NDRAT = 0) and (az mod NDRAT = 0)
            [deploys driver cell]
         End if
      Next az
   Next ay
Next ax
\end{verbatim}

%\colorbox{green}{assignment of CET values (proliferation from the zygote}
Once driver cells have been deployed, they must be assigned CET values. Figure 6 reports an example of proliferation from the zygote. To each new driver cell a new, previously unseen and unique CET value is assigned, obtained starting from the zygote's CET value (the array [0,0,0,0] in the figure, labelled with 'A') and adding 1 to the value held in the second array position at each new assignment (the second array position corresponds to step 1, in which the proliferation occurs -array counting is 0-based); with reference to the figure, the new driver cells are assigned the values [0,1,0,0],[0,2,0,0],[0,3,0,0], ... , labelled with 'B','C','D', etc. (please note that labels are just used in the figures for visualisation purposes, but all operations -essentially match-tests- are made on the underlying arrays). As far as the order of assignment is concerned, CET values are assigned to driver cells in the same order by which drivers are created.

%\colorbox{green}{assignment of CET values (proliferation from the generic driver}
In case of a proliferation triggered on a driver cell other than the zygote (figure 7), the main difference is that in this case the mother's CET value is not [0,0,0,0], but depends on the cell ``lineage'': in the case of the figure, the CET value is [0,7,4,0] (i.e. this was the 4th driver cell to be assigned in a proliferation triggered on a cell that was the 7th driver cell to be assigned in a proliferation triggered on the zygote). The rest of the procedure is unchanged: to each new driver cell a new CET value is assigned, starting from the mother cell's CET value (the array [0,7,4,0] in the figure, labelled with 'P') and adding 1 to the value of the i-th position of the array at each new assignment, where i is the current AS value (3 in the figure, corresponding to the 4th column); with reference to the figure, the new driver cells are assigned the values [0,7,4,1],[0,7,4,2], [0,7,4,3], ..., labelled with 'Q','R', 'S', etc. Therefore, apart from the mother's CET value, the only difference compared to a proliferation from the zygote is that the progressive numbers are written in the ith position of the array instead that on the 2nd position.

%\colorbox{green}{generalisations}
\color{blue} {\scshape Possible generalisations.} i) Deployment of driver cells. The exact algorithm used to deploy driver cells onto the change volume is not important, as long as two requirements are met: i) driver cells must be (more or less) uniformly distributed on the change volume (i.e. the density of driver cells has to be the same throughout the volume) otherwise some parts of the shape might end up having a lower density of driver cells, which would render them harder to develop further; ii) it has to be deterministic (driver cells must be positioned always in the same places). ii) Assignment of CET values. The CET values and the order CET values are assigned to driver cells are not important as long as two requirements are met: i) The presence of duplicated CET values must be limited (being the extent to which duplicates are present proportional to the limitation to freely explore all possible developmental trajectories); ii) the process of assignment must be subject to deterministic rules. \color{black}

\subsection{The Remove-redeploy procedure}

%\colorbox{green}{remred procedure: intro}
In case of proliferation, it may happen that the change volume is not empty. In this case the cells present in the change volume must be either deleted or moved to other positions, to make room for the cells that are going to be  created. The approach adopted, though biologically implausible, has the advantage to keep the computational burden at acceptable levels. Such approach is implemented by a procedure called 'Remove-redeploy', which intervenes in two distinct moments: before proliferation, removing the cells contained in the change volume and storing them in a temporary buffer while proliferation takes place; after proliferation is completed, redeploying the cells stored in the buffer onto free positions of the grid. Procedure details will be examined next.       

%\colorbox{green}{remred procedure: distord array}
The Remove-redeploy procedure requires the existence of an array (called 'distord' array) of triplets representing the coordinates of the points belonging to the positive quadrant of the space (the quadrant in which all coordinates have positive values) sorted according to their distance from the origin; typical choices for the distance are the euclidean distance and the Manhattan distance: in case the manhattan distance is employed the first positions of such array are occupied by the following triplets: (0,0,0), (1,0,0), (0,1,0), (0,0,1), (1,1,0), (1,0,1), (0,1,1), etc. The rule can be expressed concisely by saying that a triplet occupying a higher position in the array corresponds to a point having a greater distance from the origin. The distord array (an example of which is reported in figure 9) is the same for all proliferation events and can therefore be computed once for all at the program start. The coordinates and the quadrant are defined with reference to a 'local coordinate system', obtained by applying to the absolute coordinate system a translation that brings the mother cell in the origin and the rotation specified by the instruction's rotation matrix.

%\colorbox{green}{remred procedure: four passes}
The Remove-redeploy procedure is carried out in four passes. (1) In the first pass the change volume, the change ellissoid and the local coordinate system are drawn on the grid. (2) In the second pass, for each of the 8 space quadrants, the points of the quadrant are scanned in the order dictated by the distord array; if a cell is found which also belongs to the change volume (i.e. the cell belongs to the intersection between the quadrant and the change volume), the cell is removed from the grid and put in a temporary buffer: at the end of the scan therefore cells in the buffer are sorted according to their distance from the origin (the same distance used to compile the distord array); when the scan is completed, a pointer ptr is initialised to point to the first buffer position. (3) In the third pass the proliferation is carried out: normal and driver cells are deployed onto the change volume (which is now empty). (4) In the fourth pass, for each of the 8 space quadrants, the quadrant is scanned again, always in the order dictated by the distord array and, whenever an empty place is found, the cell in the buffer pointed by ptr is redeployed onto the grid and ptr is incremented by one; this process goes on until all cells in the buffer have been redeployed or until the boundaries of the quadrant have been reached (if the cycle is ended by this second condition, some of the cells removed will not be redeployed).

%\colorbox{green}{generalisations:push outwards}
\color{blue} {\scshape Possible generalisations.} As pointed point, the Remove-redeploy procedure is not plausible from a biological and physical viewpoint; a more realistic and plausible behaviour would be one in which the newly created cells push the existing cells outwards, which in turn would push other cells located in more external (with respect to the mother cells) positions and so forth, until the moved cells find empty positions to settle without having to displace other cells. This approach has the drawback of displacing, at each proliferation event, a significant proportion of cells and is thus very demanding from a computational viewpoint; on the other hand, from the viewpoint of the impact that the implemented procedure has on the method's effectiveness, no significant difference exists between the two approaches. \color{black}

%\colorbox{green}{generalisations:physics}
\color{blue} More generally speaking, we can say that the procedures just described are two different interpretations (at least partly) of the role of ``physics'', i.e. the set of rules by which cells are moved around and find their final position in the shape; we can add that, based on our experience, the choice of the particular physics implemented has little impact on the effectiveness of the method, as long as the deterministic requirement is maintained, i.e. as long as physics behaves in a predictable and consistent way, as we all expect. This thanks to the distribution of driver cells throughout the shape, that enables the model of development to bend any kind of physics to its goals, keeping the shape plastic during development; in fact, should (a different) physics move a driver cell away from the shape part that has to be developed, evolution would cast a change instruction on another driver, closer to the target shape part (these considerations do not refer to particularly ill-designed kinds of physics, e.g. one by which all cells are scattered in different directions, as far as possible from each other).\color{black}           

\subsection{Deployment of driver cells, generation of CET values (doping)}

%\colorbox{green}{doping: motivation}
As we have seen, proliferation events ensure that the newly created cellular mass is endowed with a uniform concentration of driver cells. On the other hand, apoptosis events (which delete cells) and repeated applications of the Remove-redeploy procedure (which move cells around the shape) could theoretically lead to a situation in which the density of driver cells is not uniform throughout the shape (some shape parts having a higher concentration of driver cells than other parts). Since driver cells are the ``foci'' that trigger the activation of change instructions, a shape region having a lower concentration of drivers would leave the evo-devo process with fewer options to choose from, with the end result of making the evolution of development harder for such region. This would in turn render some developmental trajectories less ``easy'' (and hence less likely to be taken) than other, thus creating constraints that would limit the method's effectiveness. To avoid this unwanted effect, a procedure called 'Doping' has been implemented.

%\colorbox{green}{doping: description}
The Doping procedure ``spreads'' new driver cells on the shape at the end of each age step, after the application of all change events. It works in the following way. Each normal cell C of the shape ``senses'' its neighbourhood (a sphere whose radius whose is a parameter - DOPNSZ - of the algorithm); if within such neighbourhood a driver cell is found, nothing happens; if within the neighbourhood no driver cell is found, cell C is turned into a driver cell. This procedure guarantees that the maximum distance of any normal cell from the nearest driver cell is DOPNSZ, thus ensuring that a uniform distribution of drivers is maintained throughout the shape, despite all physical events that could disrupt the even distribution of drivers provided for during proliferation events. If, on the other hand, no (major) disruption has taken place from the last run of the Doping procedure, no new drivers need to be created. In case a cell needs to be turned into a driver, as far as the generation of CET values is concerned in this case the nearest driver is taken as mother and the CET value is created by writing a digit in the array position corresponding the the current AS value (as in the case of proliferation events), taking further measures to ensure that also in this case the CET value in unique. 

%\colorbox{green}{doping:lower-level implementation}
\color{blue} {\scshape Possible generalisations.} A possible lower level mechanism to both uniformly deploy driver cells and assign CET values to them could be outlined as follows. Each existing driver cells ``emits'' a signal containing a code derived from its own CET value (and thus unique), whose strength decreases with the distance from the source (the driver itself) with a quadratic law (analogous to that governing the newtonian gravitational force); each normal cell of the shape ``senses'' all signals emitted by the surrounding drivers: if the strongest signal is below a certain threshold, the cell ``decides'' to turn itself into a driver cell. The CET value is determined based upon all signals sensed and their relevant strength (it could be composed aligning the juxtapositions of CET values and relevant strengths of all drivers whose received signal is above a given threshold); this does not guarantee 100\% that the CET values assigned are unique, but the chance the two equal CET values are created would nevertheless be rather small. 
\color{black}

\subsection{Example of development}

%\colorbox{green}{example of development}
Figure 11 shows an example of 2d development in four age steps (AS=0,1,2,3). In step 0 only the zygote is present on the grid, with CET=[0,0,0,0], whose label is 'A'. In step 1 (AS=1), this CET value triggers the activation of instruction no. 33, which has XET=[0,0,0,0] and XS=1. The instruction's right part codes for a proliferation event (ITP=0); the north-west and south-east corners' coordinates (relative to the mother's position) are (-3,-2,0) and (2,3,0) respectively (please note that, being a 2d development, all values relevant to the third coordinate are null); the rotation matrix parameters are all zero, which means that no rotation takes place; the colour code is 7, corresponding to colour pink. As a result of this proliferation event, eight new driver cells are created and evenly placed on the change volume; their CET values are:  

\begin{verbatim}
Label    CET value
 'B'     [0,1,0,0]
 'C'     [0,2,0,0]
 'D'     [0,3,0,0]
 'E'     [0,4,0,0]
 'F'     [0,5,0,0]
 'G'     [0,6,0,0]
 'H'     [0,7,0,0]
 'I'     [0,8,0,0]
\end{verbatim}

%\colorbox{green}{xxxx}
In step 2 (AS=2), two events occur. The first is caused by CET value 'D'-[0,3,0,0] triggering activation of instruction no. 9, which has XET=[0,3,0,0] and XS=2. The instruction's right part codes for an apoptosis event (ITP=1); the north-west and south-east corners' coordinates (relative to the mother's position) are (-1,-1,0) and (1,1,0) respectively; the rotation matrix parameters are all zero, which means that no rotation takes place; the colour code is 4 but, since in this event no new cells are created, it has no effect. The second event is caused by CET value 'E'-[0,4,0,0] triggering activation of instruction no. 5, which has XET=[0,4,0,0] and XS=-1. The instruction's right part codes for a proliferation event (ITP=0); the north-west and south-east corners' coordinates (relative to the mother's position) are (-1,-1,0) and (2,2,0) respectively; the rotation matrix parameters (310330000) code for an anticlockwise rotation of 45 degrees: 1 corresponds to -0.7 (=-sin(45°)), 3 corresponds to +0.7 (=sin(45°)=cos(45°)); since this is a 2d development, all other values are equal to 0; the change volume has the aspect of a ``thick diagonal'' going from south-west to north-east; the colour code is 5, corresponding to colour green. Since the change volume was not empty, the Remove-redeploy procedure has intervened, moving some cells to positions around the south-west corner of the shape. As a result of this proliferation event, two new driver cells are created and placed on the change volume; their CET values are: 

\begin{verbatim}
Label    CET value
 'J'     [0,4,1,0]
 'K'     [0,4,2,0]
\end{verbatim}

%\colorbox{green}{xxxx}
In AS=3, other two events occur. The first is caused by CET value 'F'-[0,5,0,0] triggering activation of instruction no. 23, which has XET=[0,5,0,0] and XS=3. The instruction's right part codes for a proliferation event (ITP=0); the north-west and south-east corners' coordinates (relative to the mother's position) are (-1,-4,0) and (0,0,0) respectively; the rotation matrix parameters are all zero, which means that no rotation takes place; the colour code is 6, corresponding to colour violet. The second event is caused by CET value 'I'-[0,8,0,0] triggering activation of instruction no. 37, which has XET=[0,8,0,0] and XS=3. The instruction's right part codes again for a proliferation event (ITP=0); the north-west and south-east corners' coordinates (relative to the mother's position) are (0,0,0) and (2,1,0) respectively; the rotation matrix parameters (330130000) code for a clockwise rotation of 45 degrees: 1 corresponds to -0.7 (=-sin(45°)), 3 corresponds to +0.7 (=sin(45°)=cos(45°)); since this is a 2d development, all other values are equal to 0; the colour code is '5', corresponding to colour aqua. In both events the change volume was empty, hence no intervention of 'Remove-redeploy' was needed. As a result of this proliferation event, three new driver cells are created and evenly placed on the change volume; their CET values are:

\begin{verbatim}
Label    CET value
 'L'     [0,5,0,1]
 'M'     [0,5,0,2]
 'N'     [0,8,0,1]
\end{verbatim}

\subsection{Key Features}

%\colorbox{green}{key feature 1}
In this section we would like to highlight the key features of the model of development described. The first key feature of the model is the presence of two categories of cells: normal cells and driver cells, being the latter much fewer in number (by orders of magnitude). Only driver cells have a CET value and can be instructed to develop (proliferate or die) by the Genome: they represent the scaffolding, the backbone of the developing shape and make it possible to steer the development of the whole shape by acting on a small subset of cells. If all cells (both driver and normal) had an associated CET value and would have to be guided individually by the Genome, the number of instructions needed would become unmanageable for the genetic algorithm; the fact that only driver cells are directly guided by Genome instructions makes it possible to have a number of ``bases'' in the Genome which is smaller than the total number of cells by several orders of magnitude, as it happens in nature (the human body, for instance, is made up of $\approx10^{14}$ cells, but the human genome contains ``only'' $\approx3\cdot10^{9}$ bases).

%\colorbox{green}{key feature 2}
The second feature of the model is the presence in driver cells of a variable (the CET) stored inside the cell (and moved along with the cell), that takes different values in different driver cells and represents the source of differentiation during development, leading different driver cells at different times to read out and execute different instructions in the Genome. It is by means of the cell epigenetic type that driver cells know ``who'' they are and what their behaviour has to be like; normal cells, on the other hand, do not receive guidance regarding their behaviour directly from Genome instructions but only indirectly, through driver cells. This feature represents a key difference with respect to other cellular models that rely on positional information and chemical micro-environment as basic providers of the information necessary for differentiation. In other words, in such models the differentiating information comes from the outside environment; in E.T. it comes from the outside only for normal cells, while for driver cells it comes from within the cell itself.  

%\colorbox{green}{key feature 3}
The third feature is the definition of the change events of proliferation and apoptosis in such a way that many cells (instead of one) are created/deleted at once; this increases the power of the single change event, allows a reduction of the number of change instructions needed to generate a given shape and has the end effect of speeding up considerably the morphogenetic process. Together with the previous one, this feature serves the purpose of reducing the number of the cells the Genome has to steer, at the same time endowing such cells with a capacity to influence more profoundly the course of development. From a biological perspective, proliferation events can be interpreted as ``generalised mitoses'', whose final effect is nevertheless amenable to be achieved through a series of coordinated duplications (i.e. ``standard'' mitoses): for this reason proliferation events do not appear to be in contrast with real biological processes. In the implementation described in this paper change events are only those of proliferation and apoptosis, whose effect consists in the creation and deletion of cells; more in general, change events could be defined as events originating from driver cells that ``influence'' sets of target cells in a broader sense, for instance changing some of their properties (e.g. the colour).    

%\colorbox{green}{key feature 4}
The fourth feature is represented by the mechanism of placement of new driver cells on the change volume in the course of a proliferation event, which guarantees that they are uniformly distributed in the volume, and the mechanism of assignment of the CET values on the new driver cells, which ensures that each driver cell is given a previously unseen and unique value, just as a mother gives each of her newborn babies a distinct name. This value represents the link by which these driver cells in subsequent steps can be picked up by the Genome and given other instructions to execute; if driver cells were not guaranteed to have a distinct name, the Genome would not be able to pick them individually: as a result, developmental trajectories would be biased towards certain regions of the search space, making development of arbitrary shapes harder. Moreover, thanks to the fact that the mechanism of assignment is automatic and is always the same in all proliferation events, change instructions do not have to encode also the list of CET values that will be generated (which would increase their size and make them harder to evolve). Other mechanisms (the doping procedure) are also foreseen to maintain a uniform distribution of driver cells in the face of physical events that could lead to unbalanced situations.  

\section{The evo-devo Method}

\subsection{Identification of Genome to generate a given shape}

%\colorbox{green}{evo-devo cycle}
The model of development described in the previous sections, coupled with an evolutionary technique, becomes an evo-devo method applied to the task of generating predefined 2d or 3d shapes. The method evolves a population of Genomes that guide the development of the shape starting from a small number of zygotes (usually one), for a number of generations; at each generation development is let unfold for each Genome and, at the end of it, adherence of the shape to the target shape is employed as fitness measure. Typical settings for the genetic algorithm are: population of 500 individuals (represented as strings of quaternary digits), undergoing elitism selection for up to 20000 generations, 50\% single point crossover probability and 0.1\% mutation rate per digit.
The fitness function formula is the same adopted by H. de Garis \citep{AY99DG}:
\begin{eqnarray}
\mathrm{F}=(\mathrm{ins}-\mathrm{outs})/\mathrm{des}
\end{eqnarray}
where ins is the number of cells of the evolved shape falling inside the target shape, outs is the number of cells of the evolved shape falling outside the target shape, des is total number of cells of the target shape; for coloured target shapes, also the adherence to colours is taken into account (i.e. in order to add 1 to the ins count, a given cell must fall inside the shape and its colour must be equal to that of the target cell in the same position). Figures 15-20 show some of the experiments performed. 

\subsection{Tree of CET Values}

%\colorbox{green}{Tree of CET values}
Since each CET value is created starting from a single CET value (each driver cell has only one ``mother''), the set of all CET value generated during a development has a tree-structure and therefore it is called ``Tree of CET Values'' (TCV); figures 12 shows the TCV relevant to the development of figure 11. A given CET value represents the generative history of its associated driver cell, meaning that the sequence of events that have brought to the creation of a given driver cell can be deducted from the CET value: CET value [0,4,2,0], for example, is associated to the 2nd driver cell assigned in a proliferation triggered on a driver cell that was the 4th driver cell to be assigned in a proliferation triggered on the zygote. As a consequence, the tree of CET values at the end of development contains the generative history of all CET values created during development; of course not all CET values of the tree are necessarily present on the shape at the end of development, as some may have been deleted by apoptosis events and removed from the grid.

%\colorbox{green}{junk}
An individual's TCV generated during development can be divided into i) CET values / driver cells that activate an instruction during development and ii) CET values / driver cells that {\em do not} activate any instruction during development; in the same way the individual's Genome is composed of i) XET values / instructions that become active during development and by ii) XET values / instructions that {\em do not} become active during development. By analogy with real genomes, elements in the two categories labelled with ii) can be defined as ``junk'' CET values / driver cells and ``junk'' XET values / instructions respectively (in molecular biology, ``junk DNA'' is a collective label for the portions of a genome which are never transcribed). The only measure to reduce the amount of junk in the TCV consists in decreasing the value of the parameter NDRAT (which specifies the ratio between the number of normal and driver cells generated in proliferation events): but this has the major drawback of making the distribution of driver cells sparser in the shape, weakening as a consequence the effectiveness of the evo-devo process; hence such measure cannot be undertaken and a certain amount of junk in the TCV must be reckoned with. The specular presence of junk in the Genome is caused by the procedure Germline Penetration (described later), which acts as a shuttle, transferring junk from the TCV into the Genome; in conclusion, junk in both the TCV and the Genome is a peculiar and inevitable characteristic of this method. 
  
\subsection{Germline Penetration and Progressive Freezing}

%\colorbox{green}{GP:0}
In order to develop a given shape, as we have seen, the GA has to come up with instructions whose XET value matches a CET value belonging to one of the drivers present in the shape volume. With realistic values for XET array size (greater than 2-3), the size of the space the GA has to search becomes large enough to bring the evolutionary process to a halt (if the array size is 10 and each scalar can assume 10 possible values, the search space size is $10^{10}$). A possible countermeasure would be to ``suggests'' to the GA XET values that are guaranteed or are likely to match existing CET values, instead of leaving to the GA the task of guessing them; more specifically, since we know that set of all CET values generated during development are contained in the tree of CET values, the idea is to suggest to the GA to use CET values belonging to the TCV as XET values for the instructions. This idea is implemented in a procedure called 'Germline Penetration' (figure 13), which is called at the end of each individual's development, copying at random (some) CET values from the individual's TCV (i.e. from those occurred during development) onto XET values of instructions in the Genome: such XET values will be embedded in the Genome instructions of next generation's individuals.

%\colorbox{green}{GP:1}
The name ``Germline Penetration'' draws inspiration from the imaginary path followed by the CET values present in the shape at the end of development which, from an information-flow viewpoint, leave the driver cells in which they ares tored and wander through the shape until they reach the equivalent of germline cells, which contain the genetic material that will be handed over to the subsequent generation: of course this has to be treated as a useful metaphor, as no germline cells exist in our model of development. With 'Germline Penetration' in place, the effectiveness of the evolutionary process is restored for any size of the CET array; actually, since without it the method is basically not working, this procedure should not be considered as an option, but as an integral part of the method. To avoid disrupting development, the instructions with the copied XET values are set as inactive (parameter 'ON' is set to 0): they rely on a suitable subsequent mutation affecting such parameter to become active.

%\colorbox{green}{PF:0}
Another measure to improve the effectiveness of the method is a setting called 'Progressive Freezing', which allows the evolution of an individual's Genome in blocks; the setting's formal parameters are reported in table 2.3.1 (block numbers vary from 0 to N, being k the generic block number), while an example of typical parameter values are reported in table 2.3.2. More precisely, Progressive Freezing foresees that in the course of generations from GN(k-1) to GN(k), only instructions comprised in the [XF(k),XE(k)] block are evolved, while instructions belonging to all preceding blocks are considered ``frozen'' (they cannot be changed anymore) and instructions belonging to all successive blocks are ``locked'' (they  cannot be used in the current generation block). Another feature of Progressive Freezing binds the XS fields contained in the left parts of instructions belonging to the [XF(k),XE(k)] block, to take only the value specified by XS(k). The overall effect of this setting is that age steps are evolved one by one, by a defined block of instructions and in a defined span of generations; this corresponds to implementing the ``Hackel's Hypothesis'', by which subsequent species build upon the final developmental stage of previous ones, in its purest form.

%\colorbox{green}{PF:1}
In the example of table 2.3.2, in the third stage instructions comprised in the [40-60] block are optimised, during generations in the [400-600] block and instructions' XET values are all bound to take value '2' (i.e. in this stage the whole age step 2 is evolved). Figure 14 reports the same example of development of figure 11, but evolved with the aid of Progressive Freezing. The Genome, composed of 45 instructions, is divided in blocks of 15 instructions ([0-14], [15-30] and [31-44]). Each block contains instructions which are bound to be executed only in one age step: instructions belonging to block [0-14] are all bound to be executed in age step 1, instructions belonging to block [15-30] are all bound to be executed in age step 2 and so forth (practically this means that block [0-14] instructions all have XS=1, block [15-30] instructions all have XS=2, etc.). Visually, the application of Progressive Freezing has the effect of eliminating all crossing lines from the diagram.    

\begin{verbatim}
Table 2.3.1: formal parameters
(0,  ...  ,GN(k-1), GN(k) ,GN(k+1),  ...  ,GN(N))
(0,  ...  ,XF(k-1), XF(k) ,XF(k+1),  ...  ,XF(N))
(0,  ...  ,XE(k-1), XE(k) ,XE(k+1),  ...  ,XE(N))
(0,  ...  ,XS(k-1), XS(k) ,XS(k+1),  ...  ,XS(N))
\end{verbatim}

\begin{verbatim}
Table 2.3.2: Typical parameters values
(0,  200  ,  400  ,  600  ,  800  , 1000  , 1200)
(0,    0  ,   20  ,   40  ,   60  ,   80  ,  100)
(0,   20  ,   40  ,   60  ,   80  ,  100  ,  120)
(0,    0  ,    1  ,    2  ,    3  ,    4  ,   5 )
\end{verbatim}

\section{Definitions and Abbreviations}

\begin{tabbing}
marrone \= 10 \% \kill

CET \> cellular epigenetic type \\
AS \> age step \\
ON \> instruction activation switch \\
OP \> order of precedence \\
XET \> instruction epigenetic type \\
XS \> instruction age step \\
PC \> parallelepiped corners \\
RM \> rotation matrix \\
ETP \> event type \\
COL \> colour \\
TCV \> tree of CET values \\
\end{tabbing}

\begin{verbatim}
\end{verbatim}

%\pagebreak[4]

\section{The Algorithm}

\subsection{List of Important Parameters of the Algorithm}

\begin{verbatim}
Parameter Typical value
NDIMS           3
COLOURS        16                              
GRIDX          80
GRIDY          80
GRIDZ          80
ZYGOTES        {20,20,40,20,60,40,60,20,40}    
ASMAX          18                              
CGARSZ        360                            
CETARSZ      5000
CGEVMAX        10
NDRAT           5
DOPNSZ          4
CPVMAX   30*30*30
GAGENS      20000
\end{verbatim}

%\colorbox{red}{par comments, 0}
NDIMS is the number of grid dimensions (possible values are 2 and 3).
COLOURS is the number of colours of the target.
GRIDX, GRIDY and GRIDZ are the grid sides' sizes.
ZYGOTES (array) contains the (x,y,z) coordinates of N (3 in this case) zygotes.
ASMAX is the total number of age steps foreseen.
CGARSZ is the number of instructions in the Genome.
CETARSZ is the max size of the array of CET values (max number of CET values).
CGEVMAX is the max number of events that can occur in a single age step. 
NDRAT defines the linear 'driver to normal ratio' used in proliferation events.
DOP2NS defines the linear 'driver to normal ratio' used for doping.
CPVMAX represents the maximum size (number of cells) contained in the change parallelepiped.
POPSZ the number of individuals in the genetic population
GAGENS is the number of generations.

\subsection{Pseudocode: development (Genome given)}

\begin{verbatim}
' Global variables:

Structure Lpart
   Dim on         as int ' ON value
   Dim op         as int ' OP value
   Dim xs         as int ' XS value
   Dim xet(ASMAX) as int ' XET value
End Structure

Structure Rpart
   Dim etp        as int ' event type
   Dim pc(6)      as int ' change parallelepiped corners
   Dim rm(9)      as int ' rotation matrix
   Dim col        as int ' instruction's colour
End Structure

Structure Cgi
   Dim lpart as Lpart
   Dim lpart as Rpart
End Structure

Dim cgar(CETARSZ) as Cgi  ' the array of change instructions
Dim cetar(CETARSZ)(ASMAX) as int 'is the array of CET values
Dim cr as int 'is the current number of CET values generated 
\end{verbatim}

\begin{verbatim}
Sub Devloop()

   [puts zygote(s) on the grid]
   For as = 1 To = ASMAX-1
      Cgarprep()
      Shaper()
      Doper()
   Next as

End Sub
\end{verbatim}

\begin{verbatim}
Sub Cgarprep()

    [eliminates instructions if parameter .on is 0]
    [eliminates instructions if parameter .xs is different from as] 
    [(but keeps instructions if parameter .xs is -1)] 
    [Sorts instructions in cgar according to parameter .op (ascending order)]
    ' now cgar[] is sorted and contains only instructions that can potentially fire in this as   

End Sub
\end{verbatim}

\begin{verbatim}
Sub Shaper()

   cgevnr = 0 ' change event nr
   Do While ci < cgarsz and cgevnr < CGEVXX
      ' leaves the cycle if it has browsed all instructions 
      ' or if max number of events has been reached
      ' scans the array of driver cells drvar()
      For di = 0 To CETARSZ
         If Match(cgar(ci).lpart.xet,cetar(di).cet)=YES and cgar(ci).lpart.xs=as
            ' Match is verified, pre-compiles some geometrical parameters 
            ' which will be used later
            ' (mx,my,mz) are the coordinates of the mother cell
            ' (x0,y0,z0) are the coordinates of the north-west-back corner of the c.p.
            ' (x1,y1,z1) are the coordinates of the south-east-front corner of the c.p.
            ' compiles change parallellepiped semi-sides, which coincide with change
            ' ellissoid semi-axes (esx,esy,esz)
            esx=(x1-x0)/2 : esy=(y1-y0)/2 : esz=(z1-z0)/2
            dx = {x displacement of mother cell from centre of ellissoid}
            dy = {y displacement of mother cell from centre of ellissoid}
            dz = {z displacement of mother cell from centre of ellissoid}
            [deletes mother cell] 
            If cgare(ci).rpart.etp = 0
               Removeredep(0) ' remove (necessary only in case of proliferation)
            End if
            Brush()
            cgevnr = cgevnr + 1
            If cgare(ci).rpart.etp = 0
               Removeredep(1) ' deploy (necessary only in case of proliferation)
            End if
         End if
      Next di
      
End Sub
\end{verbatim}

\begin{verbatim}
Sub Brush()

   ' compiles temporary grid buffer (buf)
   ' the centre of the ellissoid is (kx,ky,kz)
   kx = esx : ky = esy : kz = esz
   val = 0
   For ax = 0 To = 2*esx-1
      For ay = 0 To = 2*esy-1
         For az = 0 To = 2*esz-1
            ' ellissoid
            If ((pow((ax-kx)/esx,2)+pow((ay-ky)/esy,2)+pow((az-kz)/esz,2))<=1)
               If (cgar(ci).rpart.etp = 0) ' proliferation
                  [marks cell in buf(ax,ay,az) as 'normal']
               End if
               If (cgar(ci).rpart.etp = 1) ' apoptosis
                  [marks cell in buf(ax,ay,az) as 'no cell'] ' cell is deleted
               End if
               If (cgar(ci).rpart.etp = 0) and
                  (ax / NDRAT = 0) and (ay / NDRAT = 0) and (az / NDRAT = 0)
                  [marks cell in buf(ax,ay,az) as 'driver']
                  ' assigns CET value to driver cell in buf(ax,ay,az)
                  For ii = 0 To = ASMAX-1  newcet(ii) = mother.cet(ii)
                  newcet(as) = val
                  val = val + 1
                  ' assings cet to driver and updates array of drivers drvar
                  For ii = 0 To = ASMAX-1  buf(ax,ay,az).cet(ii) = newcet(ii)                  
                  For ii = 0 To = ASMAX-1  cetar(cr).cet = newcet
                  cr = cr + 1
               End if
            End if
         Next ax
      Next ay
   Next az

   'copies buffer to grid, applying translation and rotation
   For ax = 0 To = 2*esx-1
      For ay = 0 To = 2*esy-1
         For az = 0 To = 2*esz-1
            If [buf(bx,by,bz) contains a cell]
               bx = mx + dx + (rm(0)*(ax-kx) + rm(1)*(ay-ky) + rm(2)*(az-kz)
               by = my + dy + (rm(3)*(ax-kx) + rm(4)*(ay-ky) + rm(5)*(az-kz)
               bz = mz + dz + (rm(6)*(ax-kx) + rm(7)*(ay-ky) + rm(8)*(az-kz)
               If [(bx,by,bz) is inside grid]
                  grid(bx,by,bz) = buf(ax,ay,az)
               End if
            End if
         Next ax
      Next ay
   Next az
   
End Sub
\end{verbatim}

\begin{verbatim}
Sub Doper()

   For ax = 0 To = GRIDX-1
      For ay = 0 To = GRIDY-1
         For az = 0 To = GRIDZ-1
            If [cell in (ax,ay,az) is non-driver]
               drvfound = NO 'flag indicating whether a driver has been found
               For bx = ax-DOPNBSZ To = ax+DOPNBSZ
                  For by = ay-DOPNBSZ To = ay+DOPNBSZ
                     For bz = az-DOPNBSZ To = az+DOPNBSZ
                        if [cell in (bx,by,bz) is driver] drvfound = YES
                     Next bz
                  Next by
               Next bx
               If drvfound = NO
                  [determines CET value of closest driver (cetnear)]
                  [creates new CET value from cetnear]
                  [flags (ax,ay,az) as driver and gives it CET value cetnear]           
               End if
            End if
         Next az
      Next ay
   Next ax
 
End Sub
\end{verbatim}

\begin{verbatim}
Sub Removeredep(ss as int)

   ' if ss = 0 does 'remove' part, if ss = 0 does 'deploy' part
   ' distord has been pre-calculated: it contains triplets representing coordinates 
   ' of points sorted according to their distance from origin 
   ' buffer(qr) is the buffer used to store cells removed from quadrant qr
   ' ptr (qr) is a pointer of buffer(qr), incremented during 'remove' phase
   ' ptr2(qr) is a pointer of buffer(qr), incremented during 'deploy' phase
   
   For qr = 0 To = 7
      ' each assignment corresponds to a different quadrant     
      If qr = 0  ux = -1 : uy = +1 : uz = -1 ' north-west-back
      If qr = 1  ux = +1 : uy = +1 : uz = -1 ' north-east-back
      If qr = 2  ux = -1 : uy = -1 : uz = -1 ' south-west-back
      If qr = 3  ux = +1 : uy = -1 : uz = -1 ' south-east-back
      If qr = 4  ux = -1 : uy = +1 : uz = +1 ' north-west-front
      If qr = 5  ux = +1 : uy = +1 : uz = +1 ' north-east-front
      If qr = 6  ux = -1 : uy = -1 : uz = +1 ' south-west-front
      If qr = 7  ux = +1 : uy = -1 : uz = +1 ' south-east-front
      ' computes distmax as the max distance from origin of any point belonging to the
      ' change parallelepiped (i.e.) if a point has a distance from origin grater than
      ' distmax, it does not belong to the change parallelepiped. 
      distmax = max(Dist([point in change parallelepiped]))
      ' Remove
      If ss = 0 
         do
            ax = distord(ptr(qr)).xx : sx = mx + ax*ux
            ay = distord(ptr(qr)).yy : sy = my + ay*uy
            az = distord(ptr(qr)).zz : sz = mz + az*uz
            If [(sx,sy,sz) is inside the grid] and [grid(sx,sy,sz) is not empty]
               [Move cell from grid(sx,sy,sz) to buffer(qr)]
               ' computes distance of point (ax,ay,az) from origin. If such distance
               ' is grater than distmax, it means the point is does not belong to the
               ' c. p. and cycle can be quit
               dist = Dist(abs(ax),abs(ay),abs(az))               
               ptr(qr) = ptr(qr) + 1
            End if
         Loop while (dist <= distmax)
      End if
      ' Deploy
      If ss = 1 
         ptr2(qr) = 0
         For ii = 0 To = ptr(qr) ' scans buffer of removed cells
            done = NO
            do                   ' finds 1st free position
               ax = distord(ptr2(qr)).xx : sx = vx + ax*ux
               ay = distord(ptr2(qr)).yy : sy = vy + ay*uy
               az = distord(ptr2(qr)).zz : sz = vz + az*uz
               If [(sx,sy,sz) is inside the grid] and [grid(sx,sy,sz) is non empty]
                  [Move cell from buffer to grid(sx,sy,sz)] : done = YES
               End if
               dist = Dist(abs(ax),abs(ay),abs(az))
               ptr2(qr) = ptr2(qr) + 1
            while dist <= distmax and done = NO 
      End if
   Next qr
   
End Sub
 \end{verbatim}

\subsection{Pseudocode: Genome identification (target given)}

\begin{verbatim}
Sub Baseloop()

   For gn = 0 To = GAGENS 
      For [all individuals in the genetic population]
         ' development cycle
         [clears grid, puts zygote(s) on the grid]
         [decodes Genome into the cgar]
         For as = 0 To = ASMAX-1 
            Cgarprep()
            Shaper()
            Doper()
         Next as
      [Computes fitness]
      Germpenet()
      [Genetic algorithm]
   Next gn

End Sub
\end{verbatim}

\begin{verbatim}
Sub Germpenet()

   For gi = 0 To = POPSZ     'cycle on all individuals in the population
      For ci = 0 To = CGARSZ 'cycle on instructions in an individual's Genome
         [checks whether the instruction's XET matches any CET in cetar(gi)]
         If [no match is found] 
            [picks randomly a CET existing in cetar(gi)]
            rn = Rnd(10)
            If rn > 5 [copies CET onto the instruction's XET]
         End If 
      Next ci
   Next gi
   
End Sub
\end{verbatim}

%\pagebreak[4]

\begin{figure}[p] \begin{center}
{\fboxrule=0.2mm\fboxsep=0mm\fbox{\includegraphics[width=16.00cm]{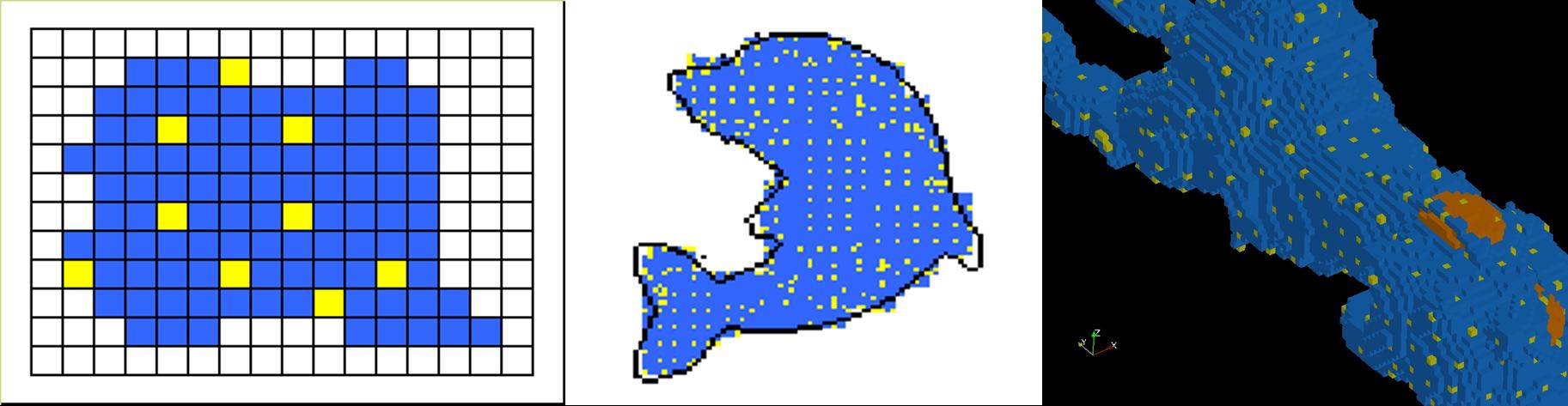}}} %[height=5.4cm] [width=8.0cm]
\caption{Examples of phenotypes in 2d and 3d.}
\label{figxx}
\end{center} \end{figure}

\begin{figure}[p] \begin{center}
{\fboxrule=0.2mm\fboxsep=0mm\fbox{\includegraphics[width=16.00cm]{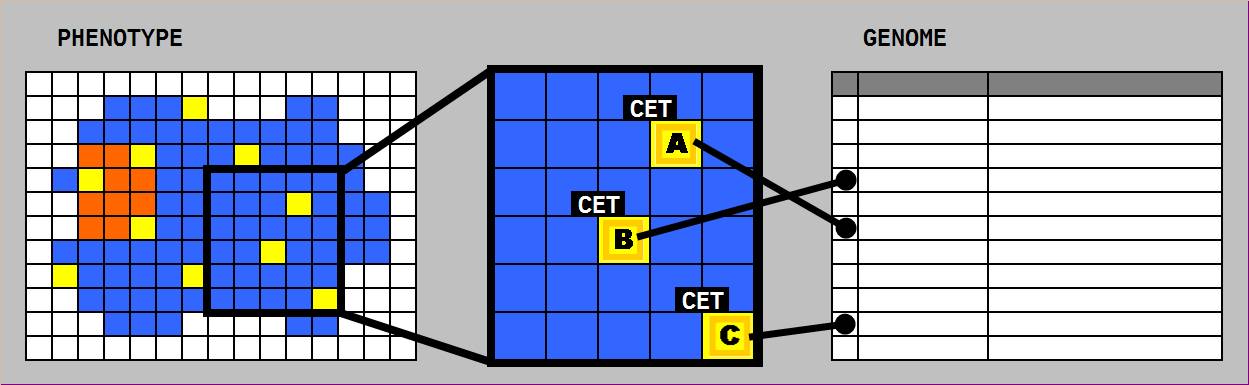}}} %[height=5.4cm] [width=8.0cm]
\caption{Driver cells, CET values and Genome.}
\label{figxx}
\end{center} \end{figure}

\begin{figure}[p] \begin{center}
{\fboxrule=0.2mm\fboxsep=0mm\fbox{\includegraphics[width=16.00cm]{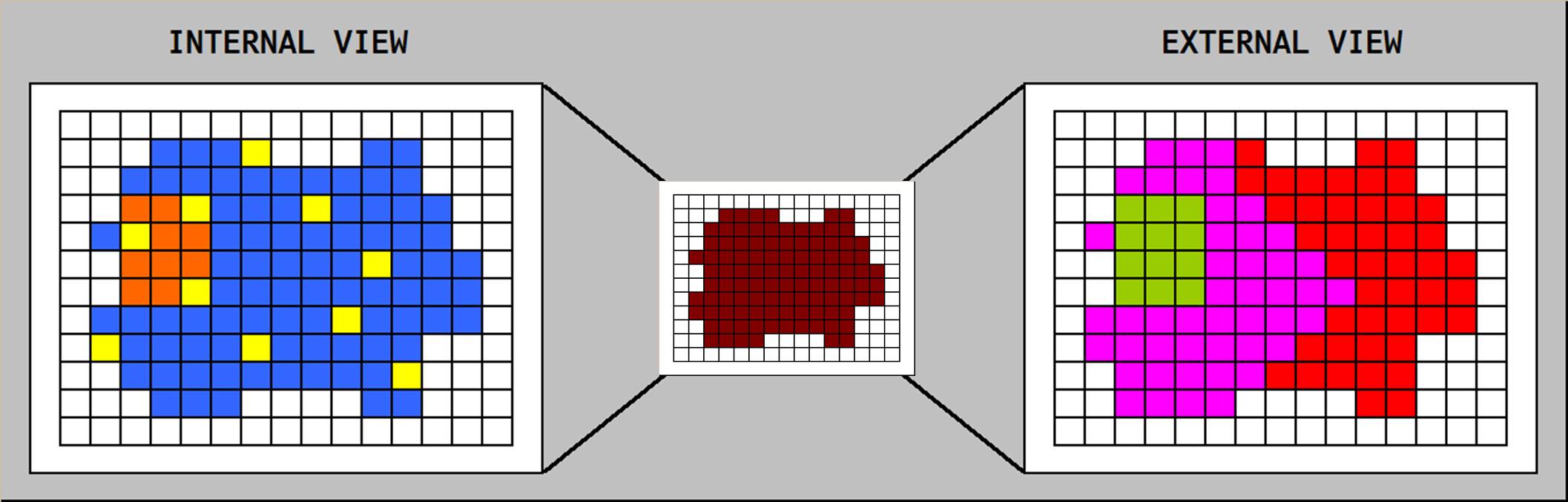}}} %[height=5.4cm] [width=8.0cm]
\caption{Internal and external view.}
\label{figxx}
\end{center} \end{figure}

\begin{figure}[p] \begin{center}
{\fboxrule=0.2mm\fboxsep=0mm\fbox{\includegraphics[width=14.00cm]{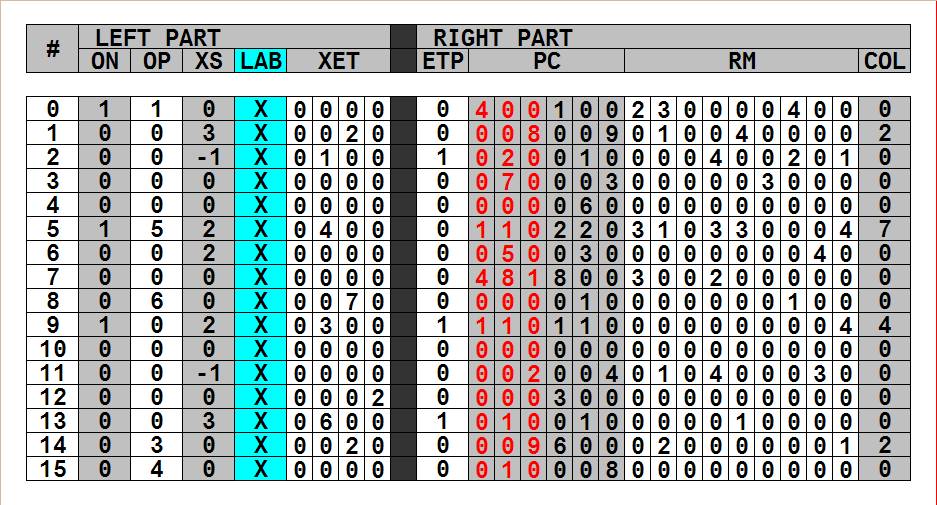}}} %[height=5.4cm] [width=8.0cm]
\caption{Example of Genome with 16 instructions (ETP = event type; PC = parallelepiped coordinates; RM = rotation matrix).}
\label{figxx}
\end{center} \end{figure}

\begin{figure}[p] \begin{center}
{\fboxrule=0.2mm\fboxsep=0mm\fbox{\includegraphics[width=18.00cm]{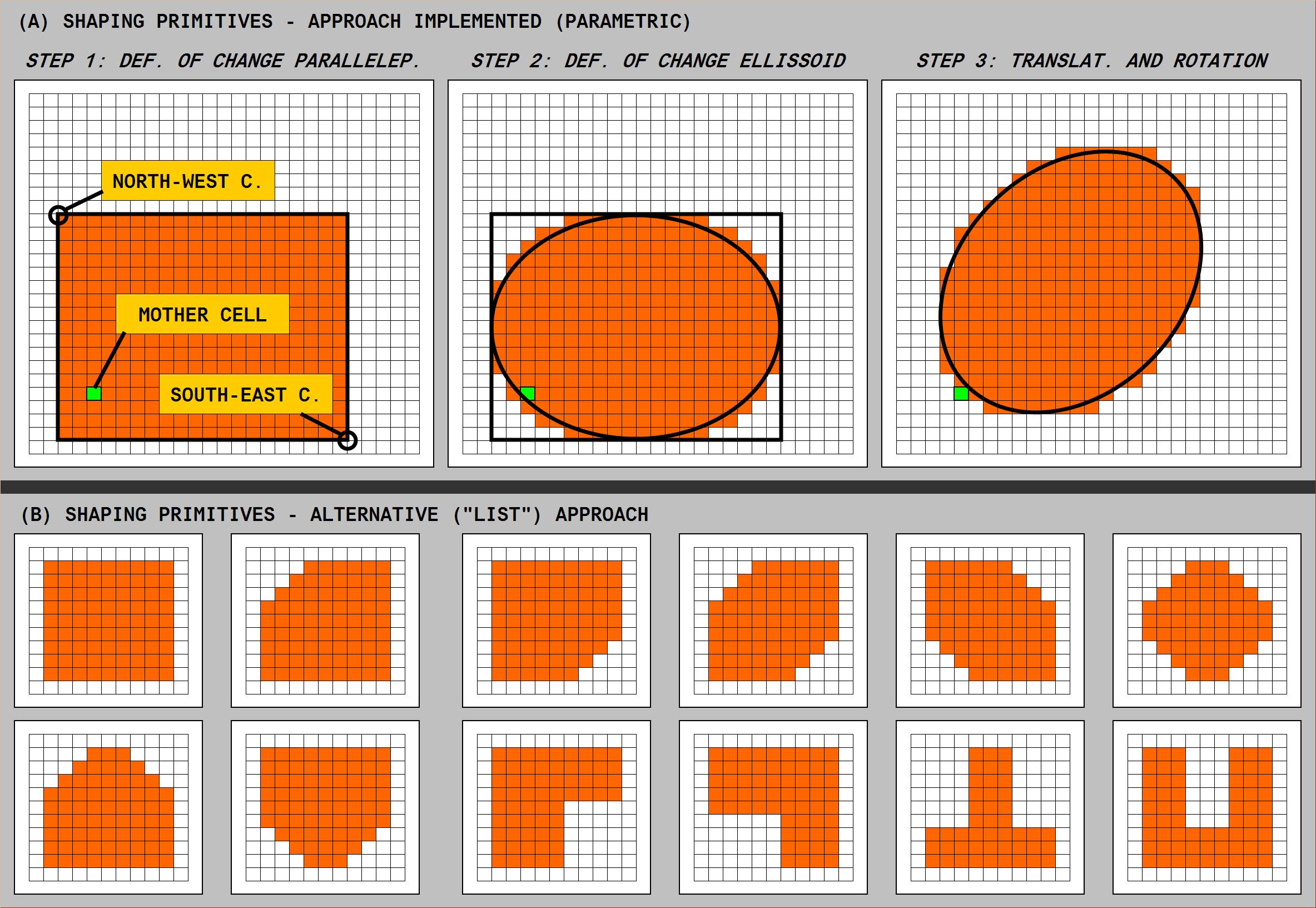}}} %[height=5.4cm] [width=8.0cm]
\caption{Definition of 2d shaping primitives (parametric and list approach).}
\label{figxx}
\end{center} \end{figure}

\begin{figure}[p] \begin{center}
{\fboxrule=0.2mm\fboxsep=0mm\fbox{\includegraphics[height=6.20cm]{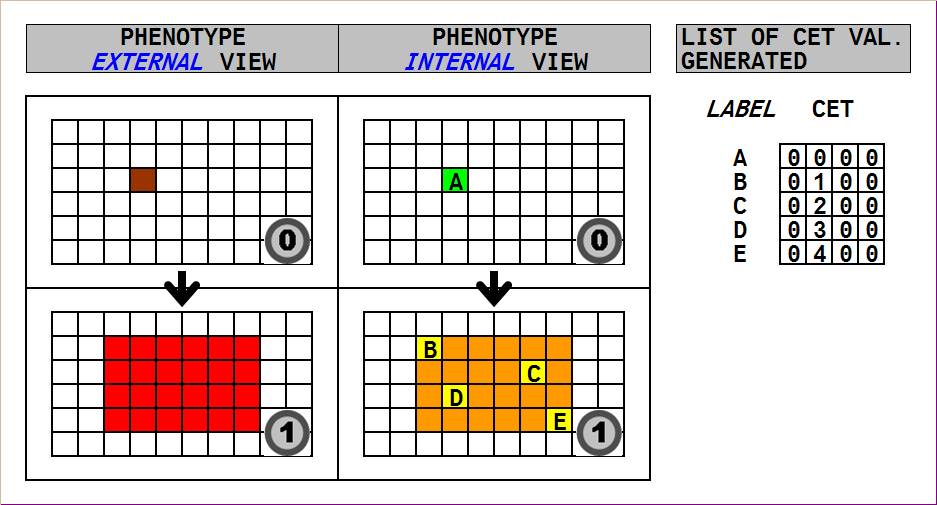}}} %[height=5.4cm] [width=8.0cm]
\caption{Proliferation triggered on the zygote (driver cells are placed manually to be roughly uniformly distributed, without reference to any particular algorithm).}
\label{figxx}
\end{center} \end{figure}

\begin{figure}[p] \begin{center}
{\fboxrule=0.2mm\fboxsep=0mm\fbox{\includegraphics[height=6.20cm]{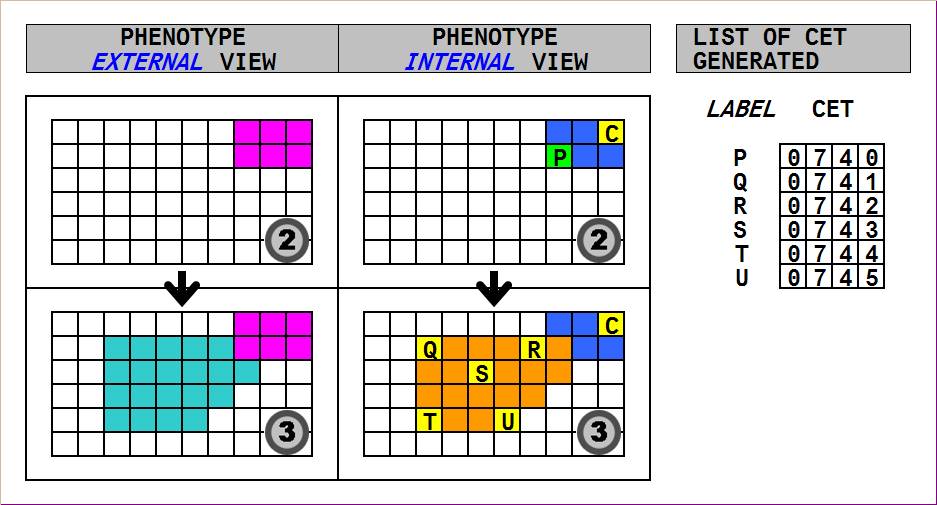}}} %[height=5.4cm] [width=8.0cm]
\caption{Proliferation triggered on a driver cell other than the zygote (driver cells are placed manually to be roughly uniformly distributed, without reference to any particular algorithm).}
\label{figxx}
\end{center} \end{figure}

\begin{figure}[p] \begin{center}
{\fboxrule=0.2mm\fboxsep=0mm\fbox{\includegraphics[height=6.20cm]{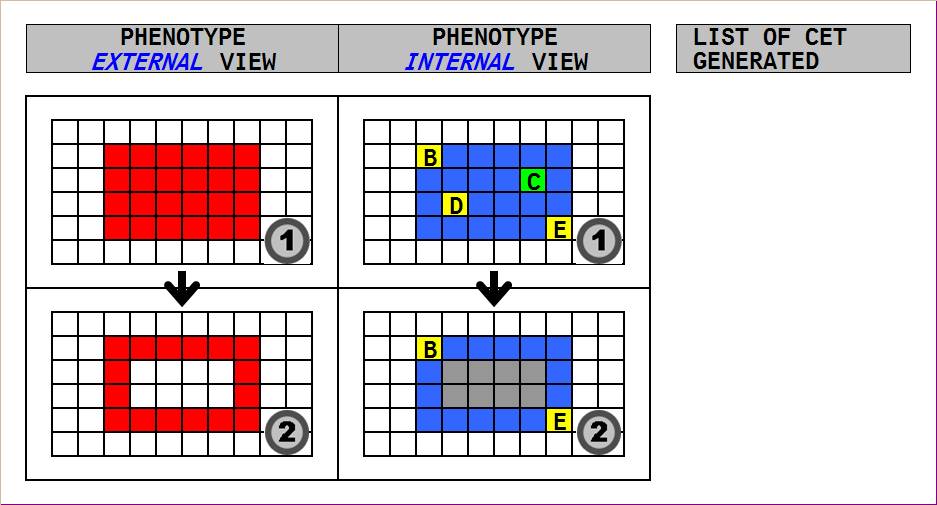}}} %[height=5.4cm] [width=8.0cm]
\caption{Apoptosis triggered on a driver cell (other than the zygote).}
\label{figxx}
\end{center} \end{figure}

\begin{figure}[p] \begin{center}
{\fboxrule=0.2mm\fboxsep=0mm\fbox{\includegraphics[width=14.00cm]{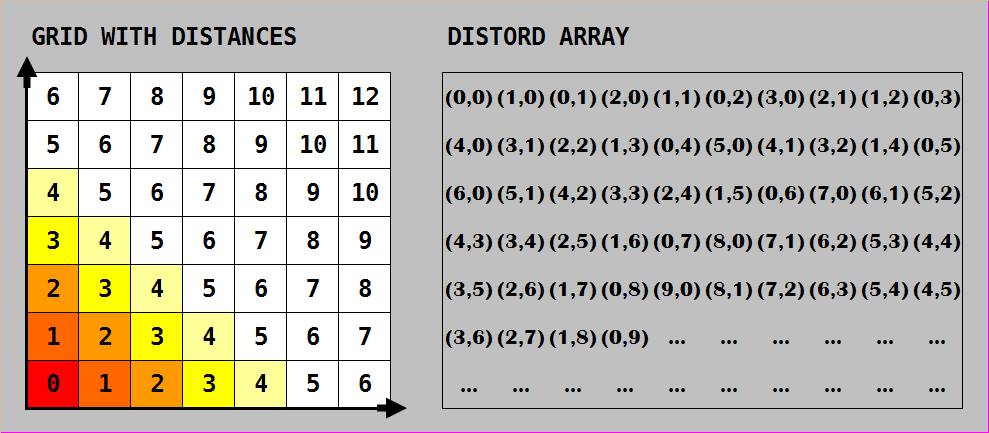}}} %[height=5.4cm] [width=8.0cm]
\caption{On the left: a 2-dimensional grid in which number represent the distance (in this case the Manhattan distance) from the origin. On the right: the distord array, containing the coordinates of grid points sorted according to their distance from the origin.}
\label{figxx}
\end{center} \end{figure}

\begin{figure}[p] \begin{center}
{\fboxrule=0.2mm\fboxsep=0mm\fbox{\includegraphics[width=14.80cm]{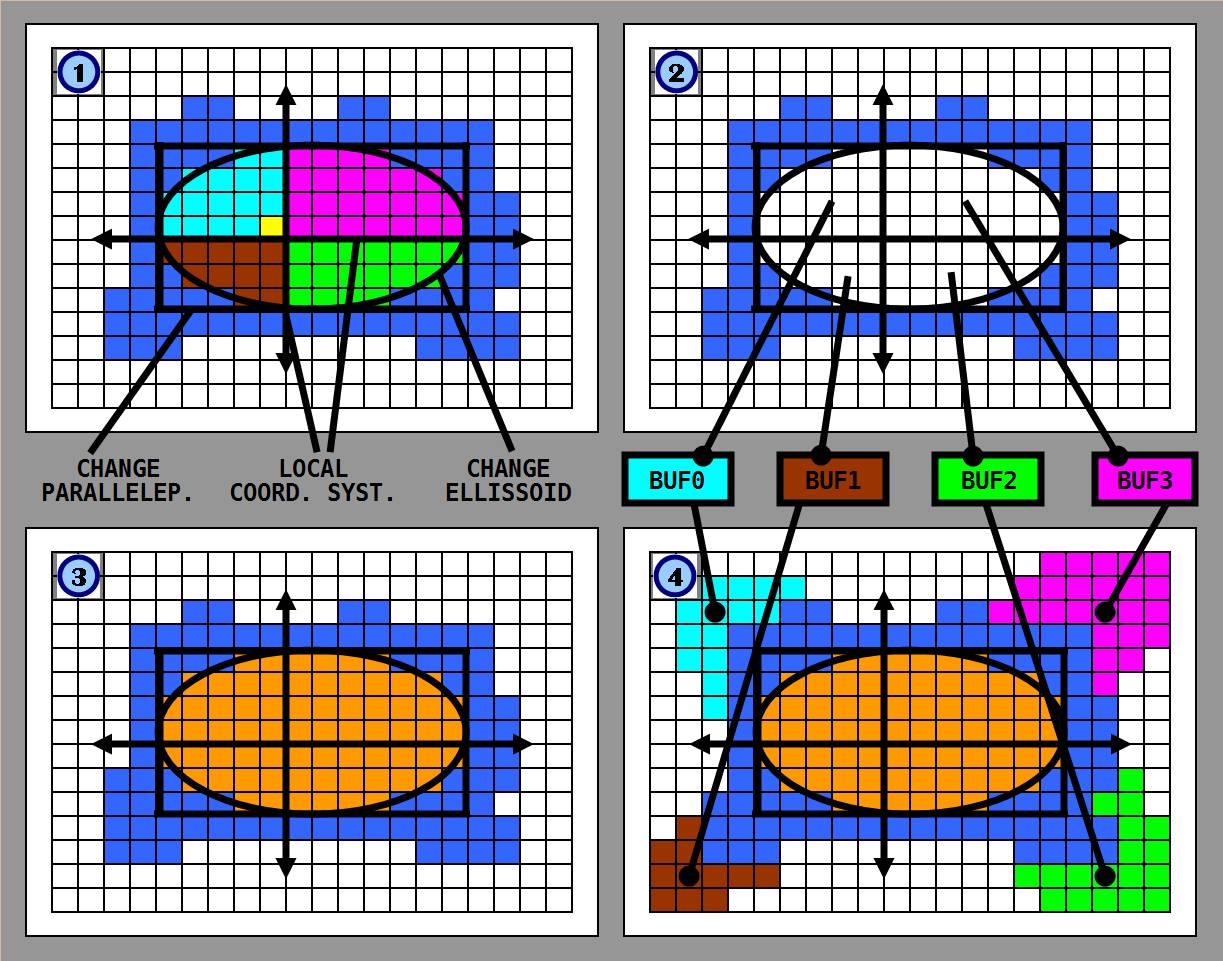}}} %[height=5.4cm] [width=8.0cm]
\caption{A schematic representation of the Remove-redeploy procedure. 1) change parallelepiped, change ellissoid and local coordinate system (with quadrants) are drawn; 2) cells belonging to the change volume are removed from the grid and put into buffers (one buffer for each quadrant); 3) proliferation is executed; 4) cells are redeployed onto the grid from buffers.}
\label{figxx}
\end{center} \end{figure}

\begin{figure}[p] \begin{center}
{\fboxrule=0.2mm\fboxsep=0mm\fbox{\includegraphics[width=18.00cm]{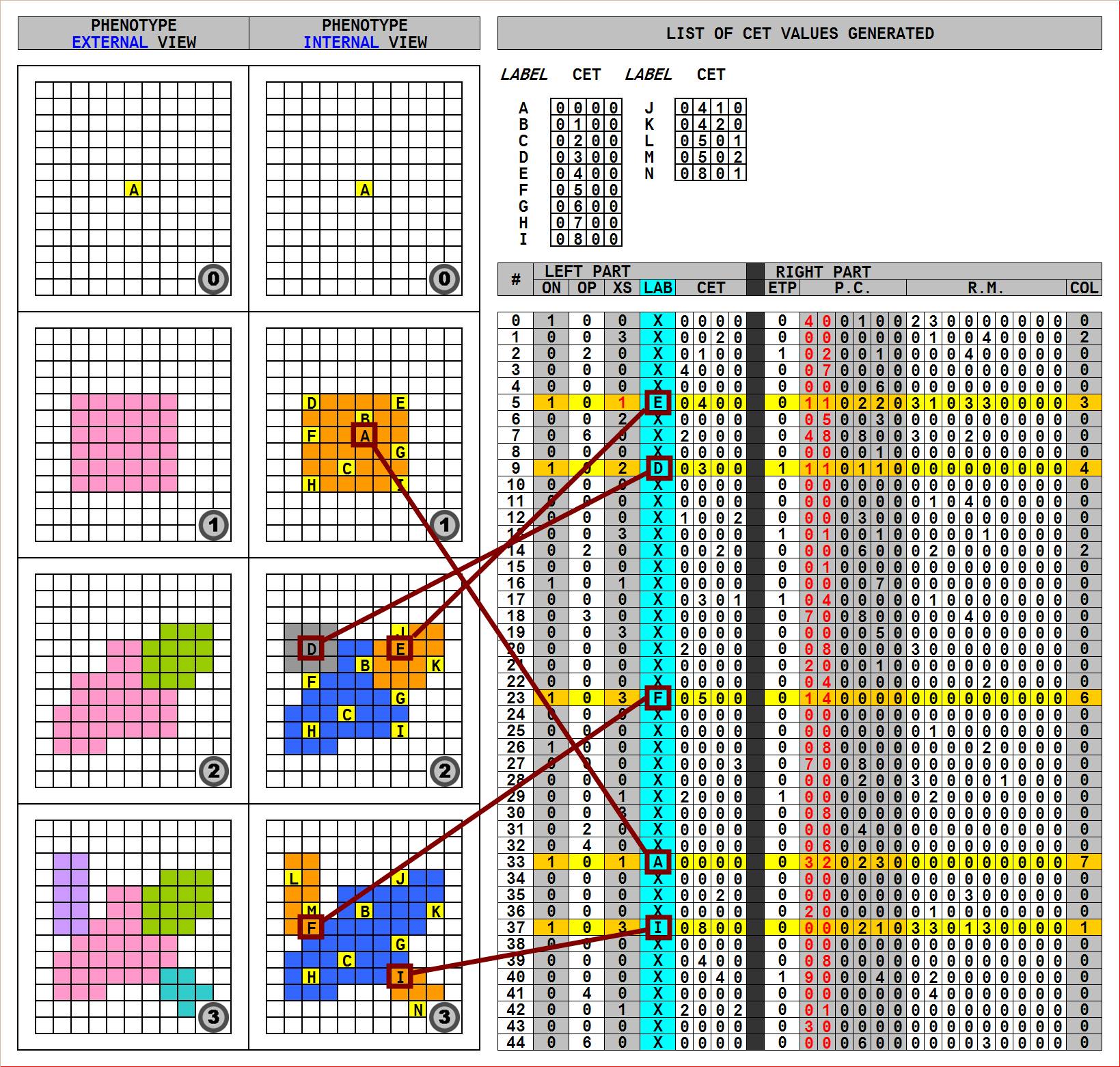}}} %[height=5.4cm] [width=8.0cm]
\caption{Example of development in four steps, driven by five instructions: Notes: i) negative numbers are in red; ii) being this an exmaple of 2d development, all numbers relevant to the third dimension (e.g. in the rotation matrix) are equal to 0; iii) in proliferation events driver cells are placed manually to be roughly uniformly distributed, without reference to any particular algorithm.}
\label{figxx}
\end{center} \end{figure}
 
\begin{figure}[p] \begin{center}
{\fboxrule=0.2mm\fboxsep=0mm\fbox{\includegraphics[width=11.00cm]{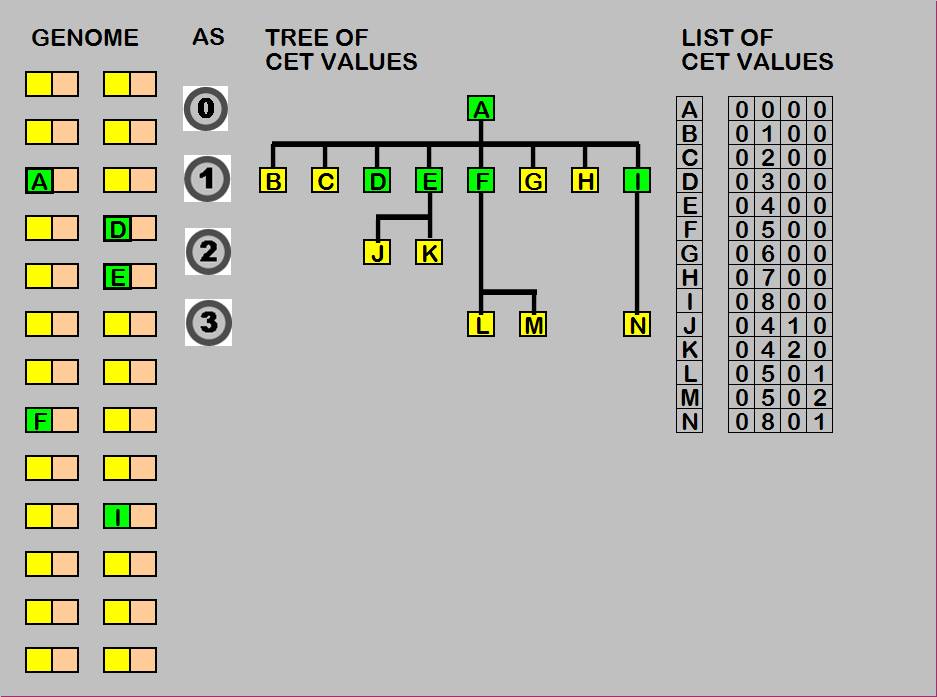}}} %[height=5.4cm] [width=8.0cm]
\caption{The Tree of CET values relevant to the development shown in figure xx. Yellow squares represent CET values that match no XET values and XET values that match no CET values (junk elements); green squares represent CET values and XET values that match.}
\label{figxx}
\end{center} \end{figure}

\begin{figure}[p] \begin{center}
{\fboxrule=0.2mm\fboxsep=0mm\fbox{\includegraphics[width=11.00cm]{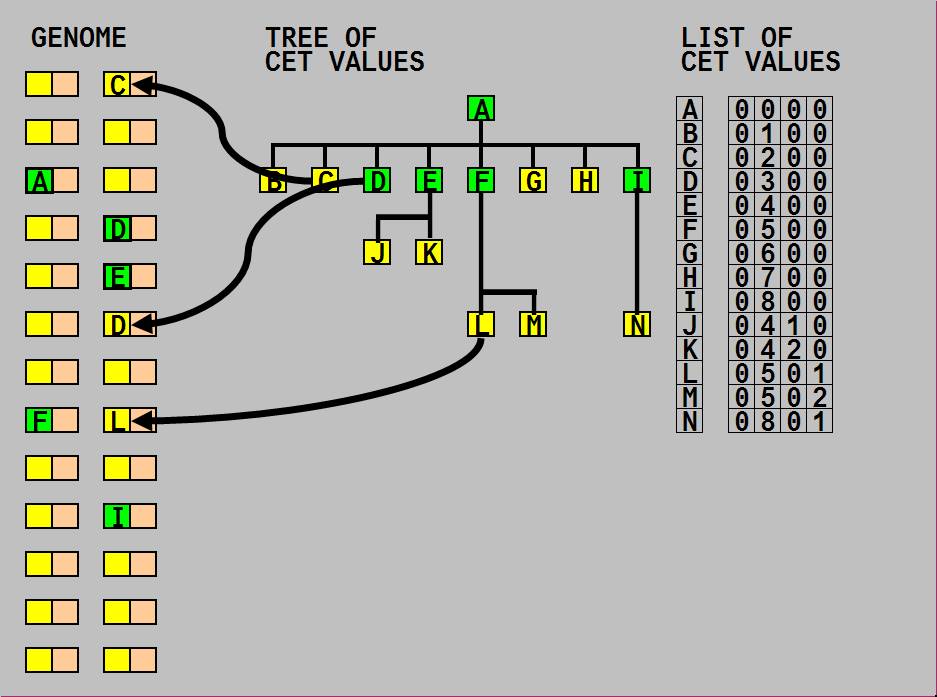}}} %[height=5.4cm] [width=8.0cm]
\caption{Germline Penetration at the end of an individual's development copies randomly CET values from the TCV into the Genome.}
\label{figxx}
\end{center} \end{figure}

\begin{figure}[p] \begin{center}
{\fboxrule=0.2mm\fboxsep=0mm\fbox{\includegraphics[width=18.00cm]{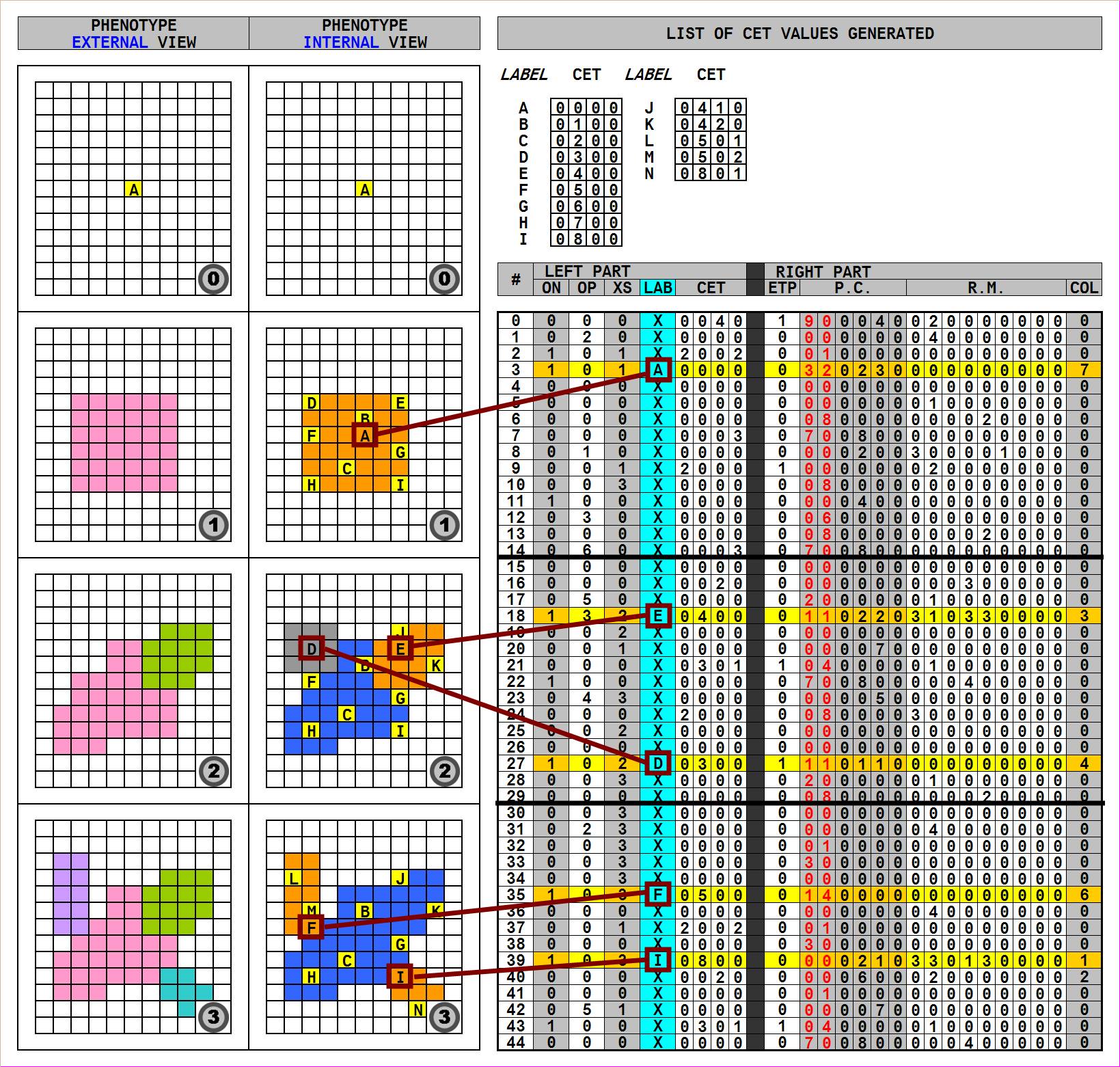}}} %[height=5.4cm] [width=8.0cm]
\caption{The same development of figure, with instructions evolved with the aid of Progressive Freezing (in three blocks of 15 instructions each).}
\label{figxx}
\end{center} \end{figure}

\begin{figure}[p] \begin{center}
{\fboxrule=0.2mm\fboxsep=0mm\fbox{\includegraphics[width=11.00cm]{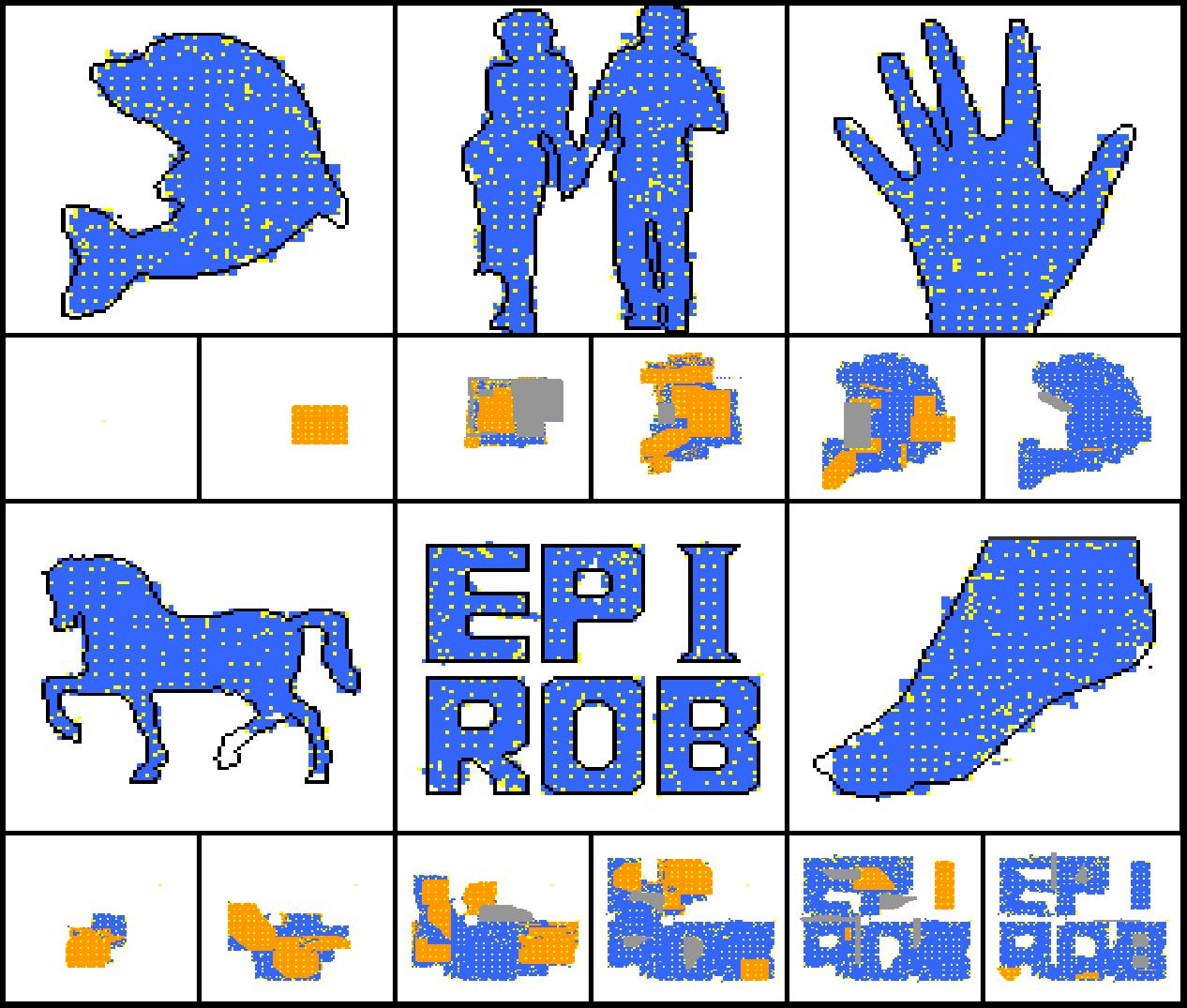}}} %[height=5.4cm] [width=8.0cm]
\caption{Experiments 2d black-and-white: the dolphin, the couple, the hand, the horse, 'EPIROB', the foot.}
\label{figxx}
\end{center} \end{figure}

\begin{figure}[p] \begin{center}
{\fboxrule=0.2mm\fboxsep=0mm\fbox{\includegraphics[width=11.00cm]{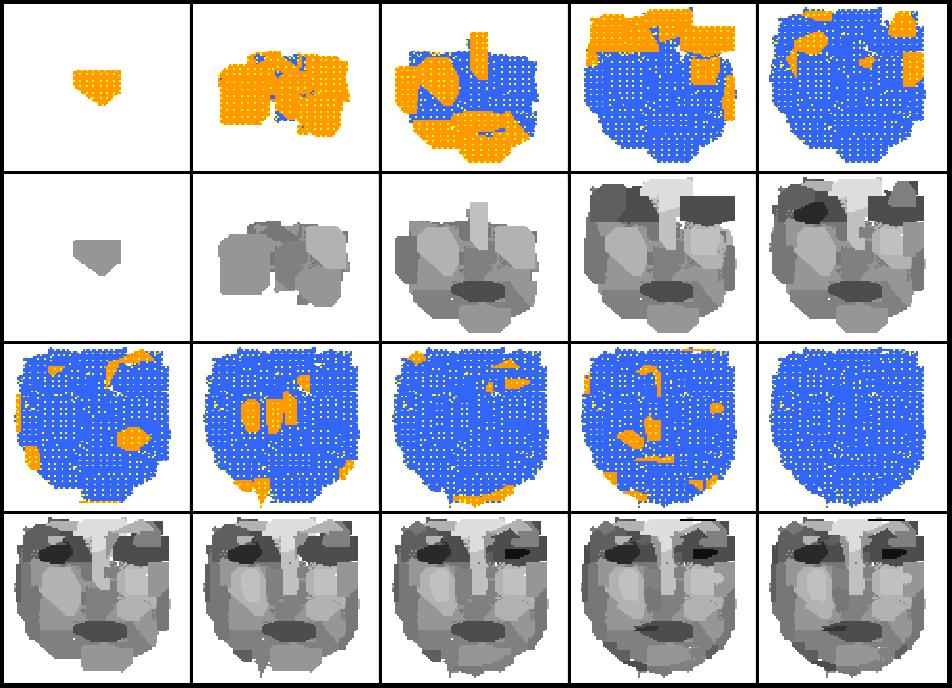}}} %[height=5.4cm] [width=8.0cm]
\caption{Experiments 2d colour: the face.}
\label{figxx}
\end{center} \end{figure}

\begin{figure}[p] \begin{center}
{\fboxrule=0.2mm\fboxsep=0mm\fbox{\includegraphics[width=10.60cm]{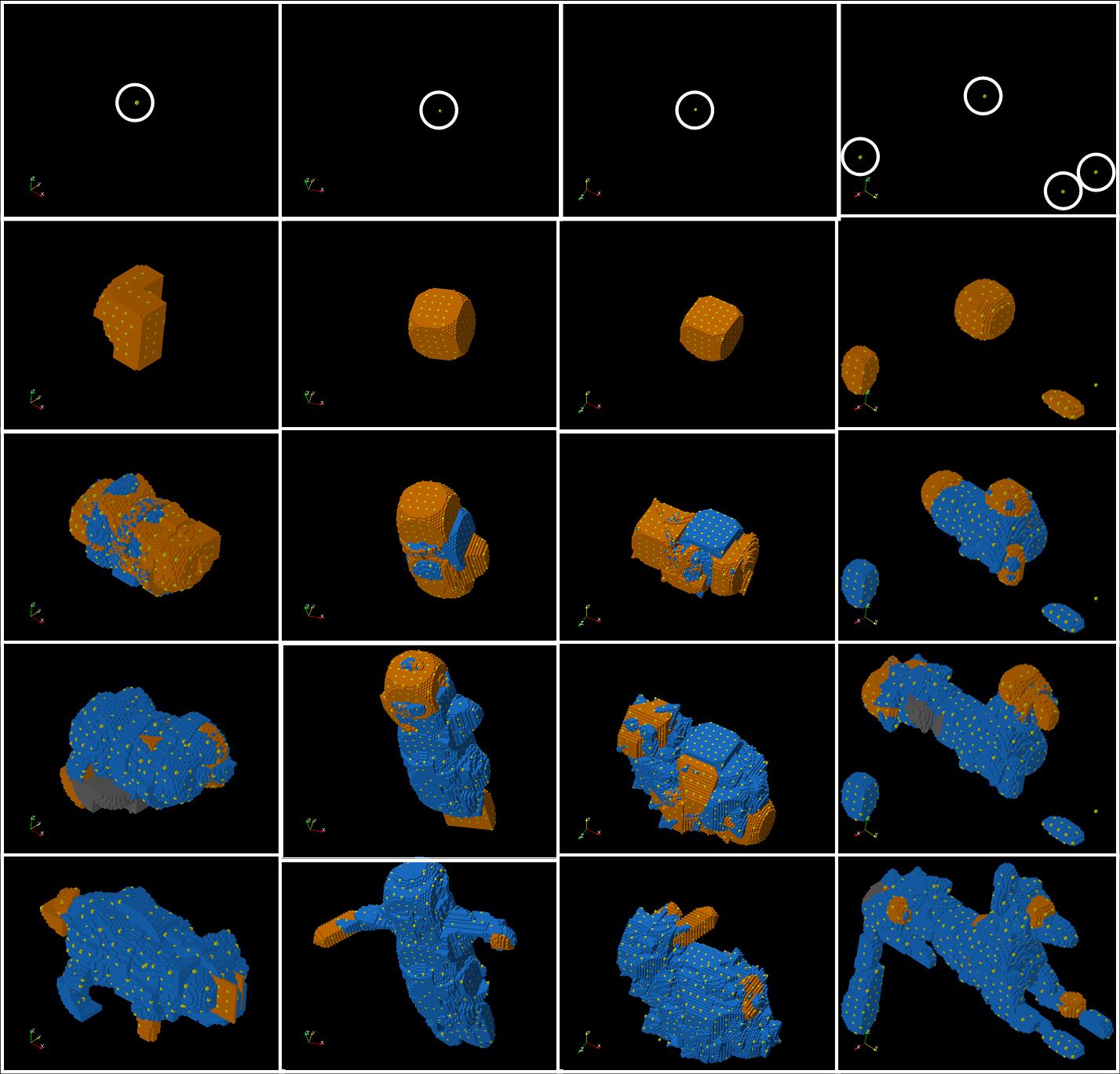}}} %[height=5.4cm] [width=8.0cm]
\caption{Experiments 3d black-and-white: the triceratops, the child, the Rufa bunny, Anubis.}
\label{figxx}
\end{center} \end{figure}

\begin{figure}[p] \begin{center}
{\fboxrule=0.2mm\fboxsep=0mm\fbox{\includegraphics[height=10.30cm]{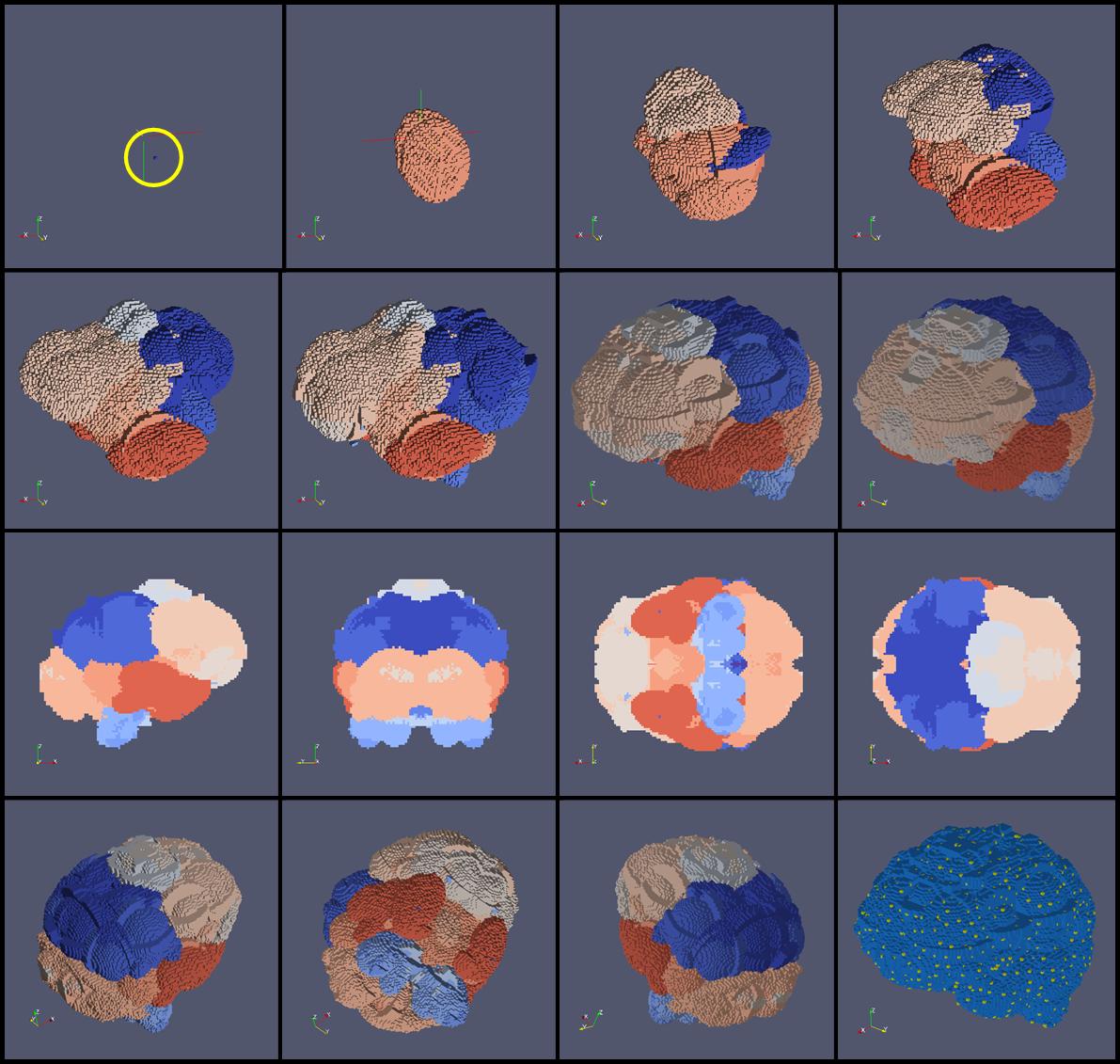}}} %[height=5.4cm] [width=8.0cm]
\caption{Experiments 3d colour: the brain.}
\label{figxx}
\end{center} \end{figure}

\clearpage

\begin{figure}[p] \begin{center}
{\fboxrule=0.2mm\fboxsep=0mm\fbox{\includegraphics[height=10.30cm]{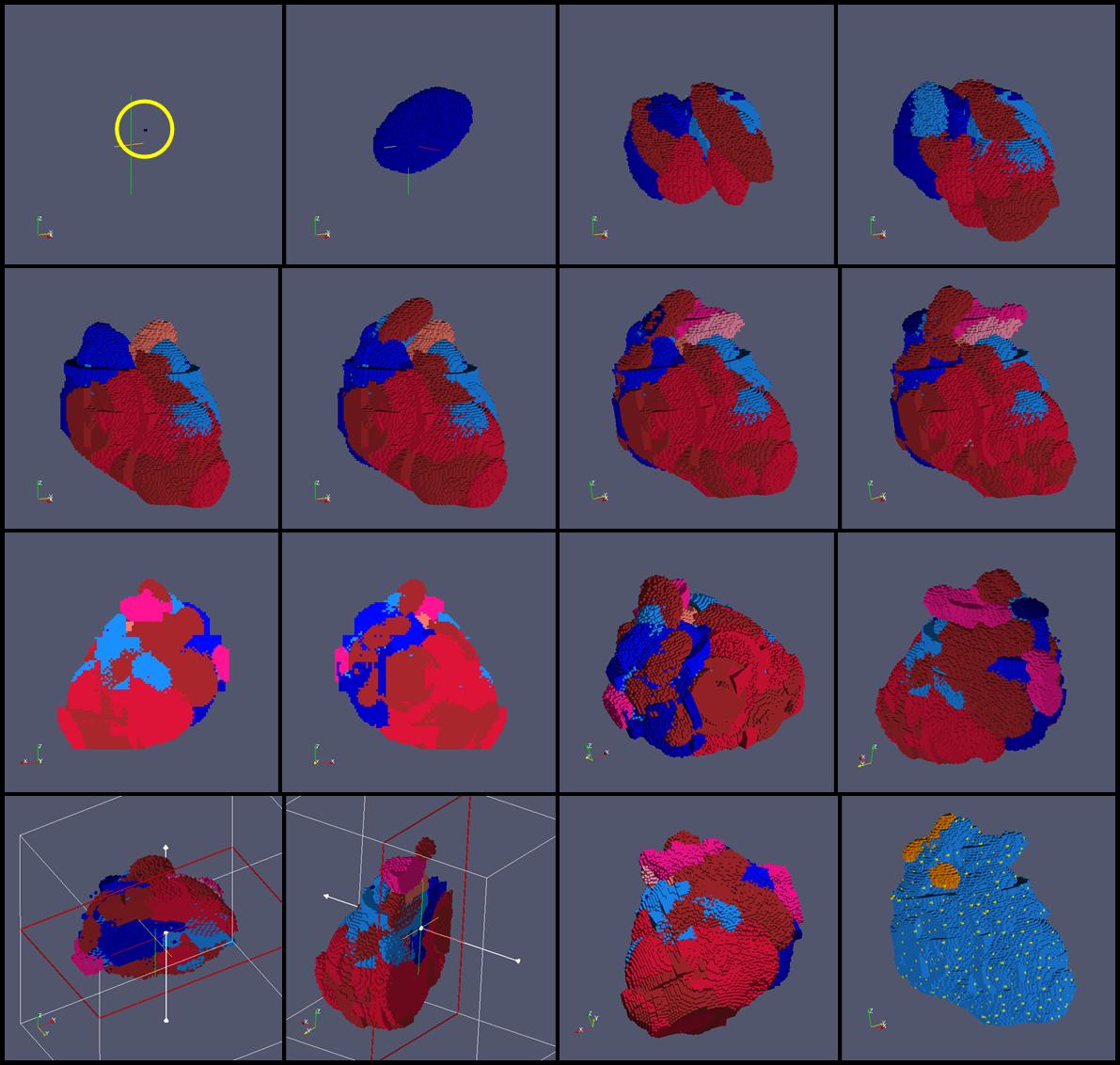}}} %[height=5.4cm] [width=8.0cm]
\caption{Experiments 3d colour: the heart.}
\label{figxx}
\end{center} \end{figure}

\begin{figure}[p] \begin{center}
{\fboxrule=0.2mm\fboxsep=0mm\fbox{\includegraphics[height=10.30cm]{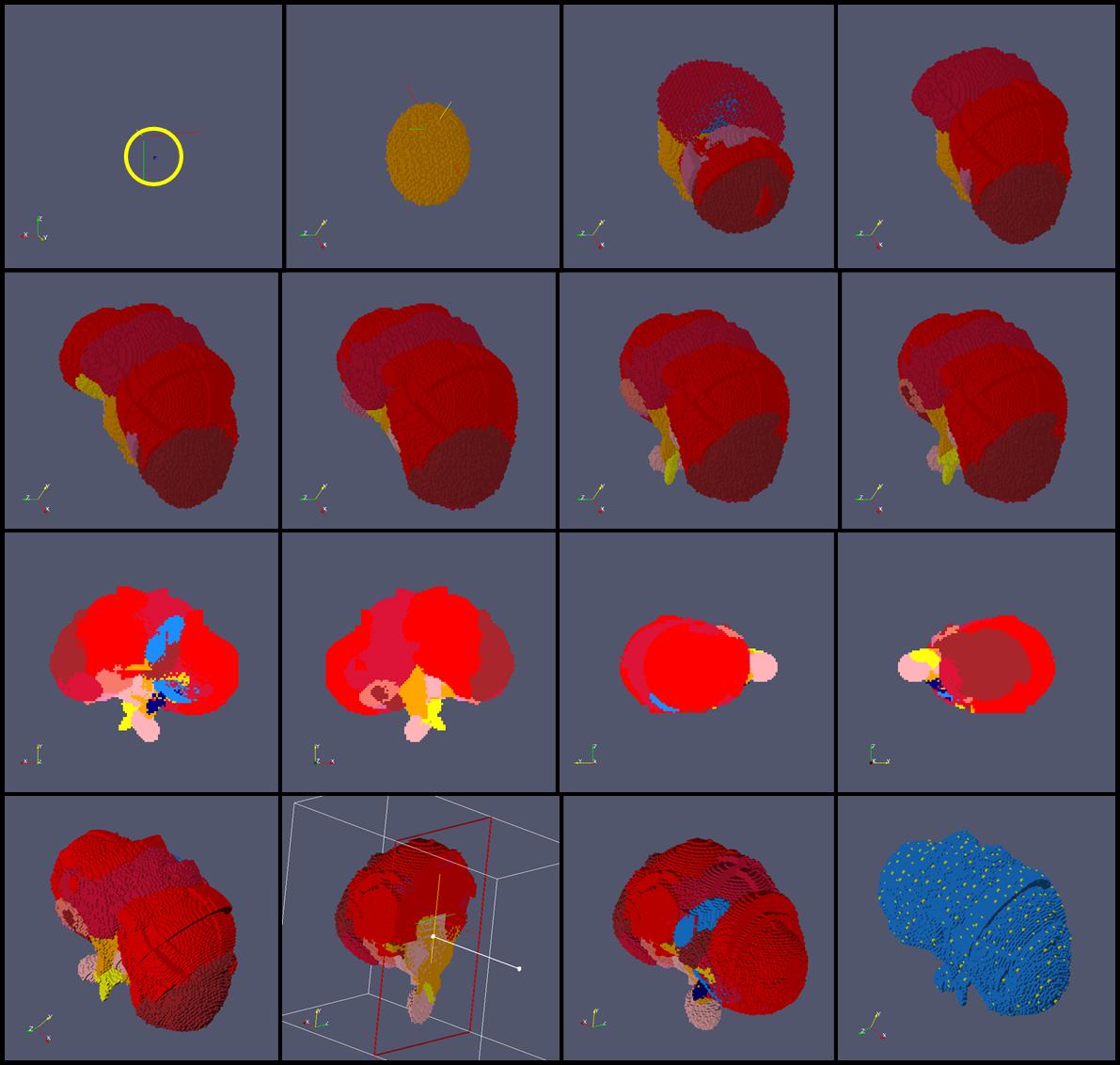}}} %[height=5.4cm] [width=8.0cm]
\caption{Experiments 3d colour: the kidney.}
\label{figxx}
\end{center} \end{figure}

\end{document}